\documentclass[a4paper,fleqn,usenatbib]{mnras}
\usepackage{graphicx}
\usepackage[T1]{fontenc}
\usepackage{ae,aecompl}

\title[The planetary nebula {\it IRAS}\,18061--2505]{The reactivation of water maser emission in the planetary nebula
  {\it IRAS}\,18061--2505 through a born-again episode}
\author[L.~F. Miranda et al. ]{L.~F. Miranda$^{1}$\thanks{E-mail: lfm@iaa.es},
  O. Su\'arez$^{2}$, L. Olgu\'{\i}n$^{3}$, R. V\'azquez$^{4}$, L. Sabin$^{4}$, P.~F. Guill\'en$^{5}$, J.~F. G\'omez$^{1}$, \newauthor L. Uscanga$^{6}$,  P. Garc\'{\i}a-Lario$^{7}$,
  I. de Gregorio-Monsalvo$^{8}$, A. Aller$^{9,10}$, A. Manchado$^{11,12,13}$, \newauthor P. Boumis$^{14}$, H. Riesgo$^{5}$, J.~M. Mat\'{\i}as$^{15}$ \\
$^{1}$Instituto de Astrof\'{\i}sica de Andaluc\'{\i}a--CSIC, C/ Glorieta de la Astronom\'{\i}a s/n, E-18008 Granada, Spain \\ 
$^{2}$Laboratoire Lagrange, UMR 7293, Universit\'e de Nice Sophia-Antipolis, CNRS, Observatoire de la C\^ote d'Azur, F-06304 Nice, \\ France \\
$^{3}$Departamento de Investigaci\'on en F\'{\i}sica, Universidad de Sonora, Blvd. Rosales Esq. L.D. Colosio, Edif. 3H, 83190 Hermosillo, \\ Son., Mexico \\
$^{4}$Instituto de Astronom\'{\i}a, Universidad Nacional Aut\'onoma de M\'exico, Apdo. Postal 877, 22800 Ensenada, B.C., Mexico \\
$^{5}$Observatorio Astron\'omico Nacional San Pedro M\'artir, Instituto de Astronom\'{\i}a, Universidad Nacional Aut\'onoma de M\'exico, 22800 \\
  Ensenada, B.C., Mexico \\
$^{6}$Departamento de Astronom\'ia, Universidad de Guanajuato, A.P. 144, 36000 Guanajuato, Gto., Mexico \\
$^{7}$European Space Agency (ESA), European, Space Astronomy Centre (ESAC), Camino Bajo del Castillo s/n, \\ 28692 Villanueva de la Ca\~nada,
  Madrid, Spain \\
$^{8}$European Southern Observatory, Alonso de Cordova 3107, Vitacura, Santiago, Chile \\
$^{9}$Departamento de Astrof\'{\i}sica, Centro de Astrobiolog\'{\i}a (INTA-CSIC), PO\,Box\,78, Villanueva de la Ca\~nada (Madrid) E-28691, Spain \\
$^{10}$Spanish Virtual Observatory, Spain \\
$^{11}$Instituto de Astrof\'{\i}sica de Canarias (IAC), E-38205 La Laguna, Tenerife, Spain \\
$^{12}$Departamento de Astrof\'{\i}sica, Universidad de La Laguna (ULL), 38206 La Laguna, Tenerife, Spain \\
$^{13}$Consejo Superior de Investigaciones Cient\'{\i}ficas (CSIC), 28006 Madrid, Spain \\
$^{14}$Institute for Astronomy, Astrophysics, Space Applications and Remote Sensing, National Observatory of Athens, \\ 15236 Penteli, Greece \\
$^{15}$Departamento de Estad\'{\i}stica, Universidad de Vigo, Campus Lagoas Marcosende, E-36200 Vigo, Spain
}
\date{Accepted XXX. Received YYY; in original form ZZZ}

\pubyear{2021}

\begin{document}
\label{firstpage}
\pagerange{\pageref{firstpage}--\pageref{lastpage}}
\maketitle

\begin{abstract}
  Water maser emitting planetary nebulae (H$_2$O-PNe) are believed to be among the youngest PNe. We present new optical narrow- and
  broad-band images, intermediate- and high-resolution long-slit spectra, and archival optical images of the H$_2$O-PN $IRAS$\,18061--2505. It 
  appears a pinched-waist bipolar PN consisting of knotty lobes with some point-symmetric regions, a bow-shock near the tip
  of each lobe, and a very compact inner nebula where five components are identified in the spectra by their kinematic and emission
  properties. The water masers most probably reside in an oxygen-rich ring tracing the equatorial region of the bipolar lobes.  
  These two structures probably result from common envelope evolution plus several bipolar and non-bipolar collimated
  outflows that have distorted the lobes. The bow-shocks could be related to a previous phase to that of common envelope.
  The inner nebula may be attributed to a late or very late thermal pulse that occurred before $\sim$1951.6 when it was not detectable in the POSS\,I-Blue
  image. Chemical abundances and other properties favour a $\sim$3--4\,M$_{\odot}$ progenitor, although if the common envelope phase accelerated
  the evolution of the central star, masses $\la$1.5\,M$_{\odot}$ cannot be discarded. The age of the bipolar lobes is incompatible with the existence
  of water masers in $IRAS$\,18061--2505, which may have been lately reactivated through shocks in the oxygen-rich
  ring, that are generated by the thermal pulse, implying that this PN is not extremely young. We discuss H$_2$O-PNe and possibly related objects in the
  light of our results for $IRAS$\,18061--2505.

\end{abstract}


\section{Introduction}

The transformation of an asymptotic giant branch (AGB) star into a planetary nebula (PN) is one of the most astonishing processes in
stellar evolution, through which most low- and intermediate-mass ($M$$\sim$0.8--8\,M$_{\odot}$) main-sequence (MS) stars 
are believed to go through at the end of their lives. An outstanding characteristic of PNe is their enormous morphological variety
(Schwarz, Corradi \& Melnik 1992; Manchado et al. 1996; Parker et al. 2006;  Sahai, Morris \& Villar 2011). Different processes are suggested
to explain their shapes, including anisotropic AGB envelopes, action of collimated outflows on the AGB envelope, and/or large-scale magnetic fields
(e.g., Sahai \& Trauger 1998; Balick \& Frank 2002; Sabin 2015; Vlemmings 2019). In a broad context, it is now accepted that 
the formation of many PNe is ultimately related to the evolution of binary/multiple
central stars (CS; De Marco 2009; Miszalski et al. 2009; Soker 2016; Jones \& Boffin 2017; and references therein).
The large parameter space of CSs with stellar, substellar, or even planetary-mass companions provides 
a context in which the varied shapes of PNe may be explained (De Marco \& Soker 2011, and references therein; Soker 2016; Reichardt et al. 2019;
Decin et al. 2020). To investigate the different situations and compare them with models, it is important to increase the number of
well-studied PNe and find possible features that could be related to binary central star evolution. 

Information about PNe formation can be obtained from the very earliest stages in this phase, in which we can study the properties of the shells
before they may be modified by further evolutionary effects. Water maser emitting PNe (hereafter H$_2$O-PNe, G\'omez et al. 2008) are believed to be among
the youngest PNe. Water maser emission, typical of oxygen-rich AGB envelopes (e.g., Engels \& Bunzel 2015), may survive $\sim$100\,yr
after strong AGB mass loss ceases (Lewis 1989; G\'omez, Mor\'an \& Rodr\'{\i}guez 1990). Therefore, stars capable to reach the PN phase
($T$$_{\rm  eff}$$\sim$25000\,K) in $\la$100\,yr from the end of AGB, are potential candidates to exhibit water masers in their very early PN
phase, implying that H$_2$O-PNe should descend from intermediate-mass MS progenitors. Nevertheless, water masers in PNe may not necessarily be
the remnants of the AGB ones but could be due to anisotropic ejections in the AGB to PN transition (e.g., G\'omez et al. 2015b).

Five bona-fide H$_2$O-PNe have been confirmed so far, using interferometric water maser observations: $IRAS$\,19255+2123
(K\,3-35, Miranda et al. 2001a), $IRAS$\,17347--3139 (de Gregorio-Monsalvo et al. 2004),
$IRAS$\,18061--2505 (G\'omez et al. 2008), $IRAS$\,16333--4807 (Uscanga et al. 2014), and $IRAS$\,15103--5754
(G\'omez et al. 2015a). There is another candidate, $IRAS$\,17393$-$2727 (G\'omez et al. 2015b), toward which water
maser emission was detected with a single-dish telescope, but no interferometric confirmation is yet available.
H$_2$O-PNe are revealing important aspects of PN formation. Interaction between jets and the AGB envelope may excite water masers
in young PNe, as suggested by Miranda et al. (2001a) to explain the water masers at the tips of the precessing jets in K\,3-35
(see also Miranda et al. 2007; Tafoya et al. 2011; Blanco et al. 2014). Magnetic fields have been detected in some H$_2$O-PNe via OH
maser polarisation (Miranda et al. 2001a; G\'omez et al. 2009, 2016; Qiao et al. 2016; Hou \& Gao 2020). Moreover, the only PN
with non-thermal (probably synchrotron) radio continuum emission identified so far is $IRAS$\,15013--5754, an extremely young, rapidly
evolving H$_2$O-PN with high-velocity water masers arising in a magnetised jet (Su\'arez et al. 2015; G\'omez et al. 2015a). 

This paper is devoted to $IRAS$\,18061--2505 (MaC\,1-10, PN\,G005.9--02.6; hereafter IRAS18061). The object was
discovered by Mac\,Connell (1978) in H$\alpha$ objective-prism plates (taken sometime between 1967 and 1974) and reported
as a possible PN. Su\'arez et al. (2006) confirmed the PN nature of IRAS18061 from a spectrum 
in the range $\sim$3600--10000\,{\AA} obtained in 1994.2. Water maser emission towards IRAS18061 was first
detected by Su\'arez, G\'omez \& Morata (2007) with single-dish observations. G\'omez et al. (2008) used the Very Large Array (VLA)
and confirmed the association between the radio continuum and water maser emission from the object.  

The H$\alpha$ image of IRAS18061 presented by Su\'arez et al. (2006) shows a bipolar nebula of $\sim$40\,arcsec in 
size with two lobes oriented at PA$\sim$60$^{\circ}$, emanating from a bright compact (stellar-like) object that is detected 
at optical, near-, mid-, far-infrared, and radio continuum wavelengths (G\'omez et al. 2008; Zhang, Hsia \& Kwok 2012) and may be better
referred to as the core of IRAS18061; the bipolar lobes have been detected so far at optical wavelengths only. The CS is hosted in the core, its
spectrum presents broad carbon emission lines, and has been classified as [WC8] by G\'orny \& Si\'odmiak (2003). G\'orny et al. (2004, 2009) reported
peculiar chemical abundances in the nebula. As usually found in PNe with late [WC]-type central stars, IRAS18061 presents dual chemistry
(Perea-Calder\'on et al. 2009; Guzm\'an-Ram\'{\i}rez et al. 2011).

IRAS18061 is a peculiar case among H$_2$O-PNe. It is the only one completely visible at optical wavelengths, including the CS, while the others are
more extincted. It is also the only H$_2$O-PN without OH maser emission (G\'omez et al. 2008), which is present in the rest (Zijlstra et al. 1989;
Sevenster et al. 1997; Miranda et al. 2001a; de Gregorio-Monsalvo et al. 2004;  Uscanga et al. 2012; G\'omez et al. 2015a,b; Quiao et al. 2016). The angular
size of the nebula ($\sim$40\,arcsec) is very large when compared with that of $\sim$2--8\,arcsec observed in the other H$_2$O-PNe. Apart from the
presence of water masers and these peculiarities, that make IRAS18061 an interesting PN, many properties of the object remain largely unknown.

In this paper we present a study of IRAS18061 based on new narrow- and broad-band optical images, high- and 
intermediate-resolution long-slit spectra, and archival optical images. After analysing the properties of IRAS18061, we 
try to estimate the mass of its MS progenitor, describe the formation process of the nebula, and propose a new 
scenario to account for the presence of water masers in this PN. Finally, we discuss H$_2$O-PNe and possibly related
objects in the light of the results obtained for IRAS18061 and evolutionary models for the AGB to PN transition. 

\section[]{Observations}

\subsection{Optical imaging}

Narrow-band optical images of IRAS18061 were obtained on (1) 2000 June 
12 with ALFOSC at the 2.5\,m Nordic Optical Telescope (NOT) at El Roque de los Muchachos Observatory 
(La Palma, Spain); (2) 2010 July 30 with CAFOS at the
2.2\,m telescope at Calar Alto Observatory (Almer\'{\i}a, Spain), and (3) 2013 August 7--9 
with the 2.3\,m Aristarchos Telescope at Helmos Observatory of the National Observatory of 
Athens (Greece). Broad-band images were obtained with CAFOS on 2017 June 28.

(1) The detector of ALFOSC was a Tektronik 1k$\times$1k CCD with a plate scale of 
0.176\,arcsec\,pixel$^{-1}$. Images were obtained through the H$\alpha$ ($\lambda$$_0$ = 6563\,{\AA }, FWHM = 9\,{\AA }), 
[N\,{\sc ii}] ($\lambda$$_0$ = 6584\,{\AA }, FWHM = 9\,{\AA }), and [S\,{\sc ii}] ($\lambda$$_0$ = 6725\,{\AA },  
FWHM = 60\,{\AA }) IAC filters, and through the [O\,{\sc iii}] ($\lambda$$_0$ =
5007\,{\AA }, FWHM = 30\,{\AA }) NOT filter. 
Exposure time was 300\,s in the H$\alpha$, [N\,{\sc ii}], and [O\,{\sc iii}] filters, and 600\,s in the 
[S\,{\sc ii}] one. Seeing was $\sim$1.1\,arcsec. 

(2) The detector of CAFOS was a SiTe 2k$\times$2k CCD with a plate scale of 0.53\,arcsec\,pixel$^{-1}$. 
Images were obtained through H$\alpha$ ($\lambda$$_0$ = 6563\,{\AA}, FWHM = 15\,{\AA}) and 
[O\,{\sc iii}] ($\lambda$$_0$ = 5007\,{\AA }, FWHM = 87\,{\AA })
filters. Exposure time was 1800\,s in each filter. Seeing was
$\sim$1.4\,arcsec. The broad-band images were obtained 
in the Johnson B, V, R, and I filters with an exposure time of 900\,s in each
filter, poor seeing conditions ($\sim$3.5\,arcsec) and non-photometric night.  

(3) At the Aristarchos telescope, the detector was an E2V 1k$\times$1k CCD with a plate scale of 
0.28\,arcsec\,pixel$^{-1}$. Images were obtained through H$\alpha$ ($\lambda$$_0$ = 6567\,{\AA }, FWHM = 17\,{\AA }), 
[N\,{\sc ii}] ($\lambda$$_0$ = 6588\,{\AA }, FWHM = 17\,{\AA }), and [O\,{\sc iii}] ($\lambda$$_0$ = 5011\,{\AA },  
FWHM = 30\,{\AA }) filters. Exposure time was 1800\,s in each filter. Seeing was $\sim$1.5\,arcsec.

The images were cosmic rays cleaned, bias subtracted, and flat fielded using 
standard procedures in the {\sc midas} package.

\begin{figure*}
\begin{center}
\includegraphics[width=160mm]{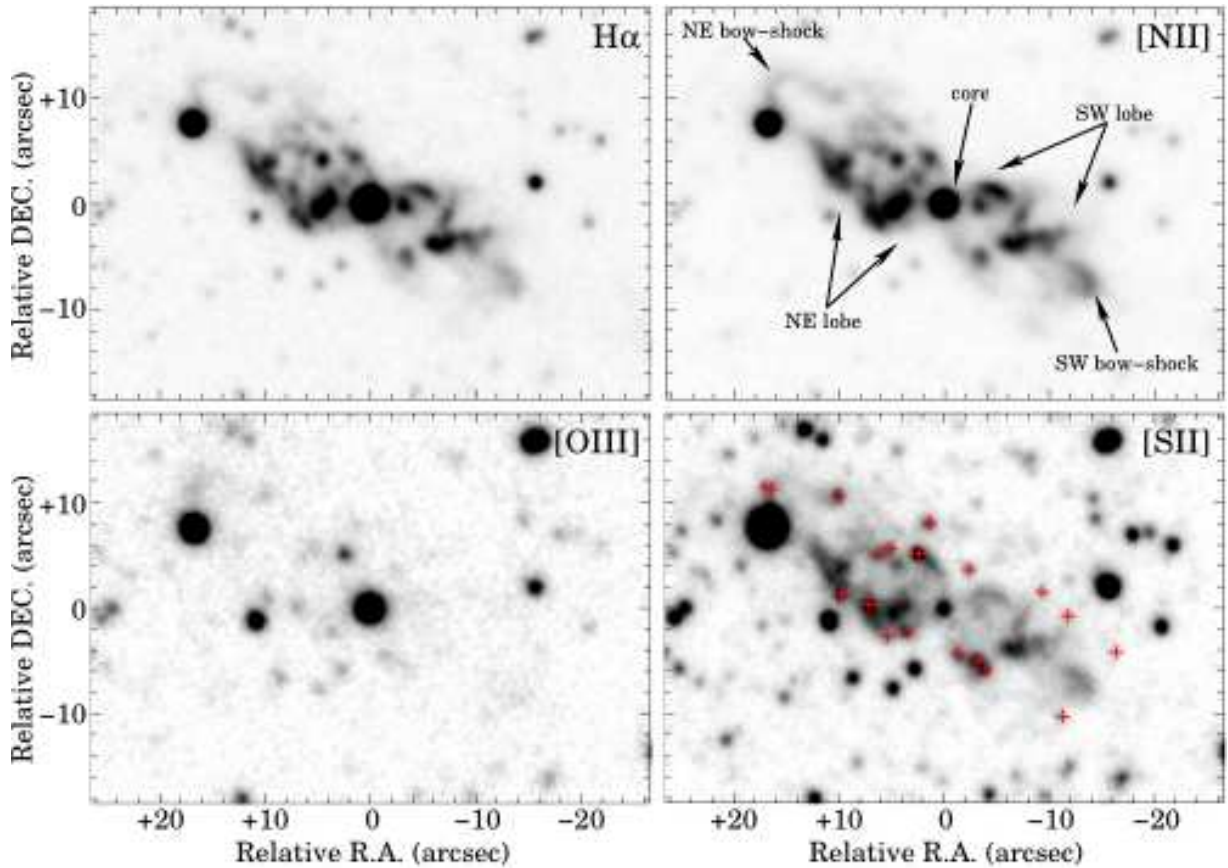}
\caption{Grey-scale reproductions of the H$\alpha$, [N\,{\sc ii}], [O\,{\sc iii}], and [S\,{\sc ii}] NOT images of IRAS18061. 
The grey levels are linear. The origin (0,0) is located at the position of the maximum intensity of the 
core at $\alpha$(2016.0) = $18^{\rm h}$ $09^{\rm m}$ $12\rlap.^{\rm s}411$, $\delta$(2016.0) = $-25^{\circ}$ 04$'$ $34\rlap.^{''}56$
from the Gaia Early Data Release\,3. The basic nebular structures discussed in the paper are marked in
the [N\,{\sc ii}] panel. Red crosses in the [S\,{\sc ii}] image mark field stars on or close the nebula, as deduced from
the PanSTARRs r and z images (see Appendix\,A). Seeing is $\sim$1.1\,arcsec.}
\end{center}
\end{figure*}

\subsection{High-resolution long-slit spectroscopy}

High-resolution, long-slit spectra were obtained with the Manchester Echelle
Spectrometer (MES, Meaburn et al. 2003) at the 2.12\,m telescope at the San Pedro M\'artir Observatory (OAN-SPM, Baja California, 
M\'exico) on 2008 June 4--6, 2015 May 16, and 2017 August 7. In 2008 the detector was a SiTE
1k$\times$1k CCD (in 1$\times$1 binning), providing wavelength and angular scales of 0.05\,{\AA }\,pixel$^{-1}$ and 
0.33\,arcsec\,pixel$^{-1}$, respectively. In 2015 and 2017 the detector was an E2V
2k$\times$2k CCD (in 2$\times$2 binning), providing wavelength and angular scales of 0.057\,{\AA }\,pixel$^{-1}$ and
0.351\,arcsec\,pixel$^{-1}$, respectively. MES has no cross-dispersion, hence, a $\Delta$$\lambda$=90\,{\AA } band-width
filter was used to isolate the 87$^{\rm th}$ order covering the H$\alpha$ and [N\,{\sc  ii}]$\lambda$6583 emission lines in the 
2008 spectra and also including the [N\,{\sc  ii}]$\lambda$6548 emission line in the 2015 {\bf and 2017} spectra. The slit, with
a length of 6.5\,arcmin and a width of 1(2)\,arcsec in the 2008(2015,2017) observations, was centred on the core of
IRAS18061 and oriented at position angles (PAs) +55$^{\circ}$, +26$^{\circ}$, --35$^{\circ}$, and 
--81$^{\circ}$ in 2008, PA +55$^{\circ}$ in 2015 (hereafter PA +55$^{\circ}$/2015), and PAs +33$^{\circ}$ and +60$^{\circ}$ in 2017.
We note that the 2008 spectra have a lower signal-to-noise ratio than that those obtained in 2015 {\bf and 2017}. Exposure time was 1800\,s for
each spectrum. Seeing was $\sim$2\,arcsec in the three epochs.

The spectra were reduced using standard procedures for long-slit spectroscopy in the {\sc
  iraf}\footnote{{\sc iraf} is distributed by the National Optical Astronomy Observatory, which is operated by the 
Association of Universities for Research in Astronomy (AURA) under a cooperative agreement with the National 
Science Foundation.} package. Wavelength calibration was carried out using a Th-Ar lamp and the spectra 
were calibrated to an accuracy of $\pm$1\,km\,s$^{-1}$. The spectral resolution is 
$\sim$12\,km\,s$^{-1}$, as indicated by the FWHM of the Th-Ar lamp emission
lines.

\subsection{Intermediate-resolution long-slit spectroscopy}

Intermediate-resolution, long slit-spectra were obtained on 2011 July 7 with the Boller \& Chivens spectrograph mounted at the 2.12\,m telescope at the
OAN-SPM, using as detector a Marconi 2k$\times$2k CCD. We employed a 400\,lines\,mm$^{-1}$ dispersion grating along with a 2.5\,arcsec slit width,
giving a spectral resolution (FWHM) of $\sim$8.5\,{\AA} and covering the 4200--7600\,{\AA} spectral range. The slit was centred 
on the core of IRAS18061 and oriented at PA +55$^{\circ}$. Two 1800\,s exposures were obtained, reduced independently, and added to get a final spectrum
with a total exposure time of 3600\,s. Spectrophotometric standard stars were observed in the same night as the object for flux calibration. Seeing
was $\sim$2\,arcsec during the observations. Spectra reduction was carried out following standard procedures in {\sc xvista}\footnote{XVISTA was originally
developed as Lick Observatory Vista. It is currently maintained by Jon Holtzman at New Mexico State University and is available at
http://ganymede.nmsu.edu/holtz/xvista.}.

\section{Results}

\subsection{Morphology}

The H$\alpha$, [N\,{\sc ii}], [O\,{\sc iii}], and [S\,{\sc ii}] NOT images of IRAS18061 are shown in Figure\,1 in a linear
flux scale. Figure\,2 presents the [N\,{\sc ii}] NOT image in a logarithmic flux scale to better show the faint nebular regions. 
Figure\,3 presents a colour composite image obtained by combining the H$\alpha$, [N\,{\sc ii}], and [O\,{\sc iii}] images shown in
Figure\,1. To construct the colour image, we have carried out an approximate flux calibration of the individual ones using the observed
fluxes of the three  emission lines in the intermediate-resolution long-slit spectrum (Section\,3.3). 

IRAS18061 appears as a pinched-waist, knotty bipolar PN with some point-symmetric regions and polar low-ionization features.
The images allow us to identify three basic nebular structures (Figure\,1): (1) the bright core at the centre of the object; (2) 
the bipolar lobes; and (3) two bow-shock-like structures at the tips of the lobes. The long-exposure Aristarchos and CAHA images do not
show additional structures when compared with the short-exposure NOT images. In the following we will discuss the three components separately. 

\subsubsection{The core}

The core is bright in H$\alpha$, [N\,{\sc ii}], and [O\,{\sc iii}], and noticeably weaker in 
[S\,{\sc ii}] (Figure\,1). Although it appears ``stellar-like'', we noticed that 
its FWHM is larger than that of the field stars in each image, indicating that it is partially resolved. We have obtained the
PSF of each image from several field stars of similar brightness to that of the core, which was used to derive the deconvolved
size (FWHM) of the core in each image. Averaging the results in each filter, the deconvolved (FWHM) size of the core is   
0.34$\pm$0.14\,arcsec in H$\alpha$, 0.45$\pm$0.10\,arcsec in [N\,{\sc ii}], 0.70$\pm$0.15\,arcsec in [O\,{\sc iii}],
and 0.28$\pm$0.14\,arcsec in [S\,{\sc ii}]. Moreover, from the radio continuum data presented by G\'omez et al. (2008) and assuming 
that the emission at 1.665\,GHz, detected only from the core, is optically thick, a (FWHM) size of $\sim$0.41\,arcsec is obtained. These results
reveal the existence of a very (angularly) small ionised nebula in the core, that will be referred hereafter to as the inner nebula. Its deconvolved
size is maximum in [O\,{\sc iii}], minimum in [S\,{\sc ii}], and presents intermediate values in [N\,{\sc ii}] and H$\alpha$. This is not expected in
a simple photoionised nebula, where high excitation should be observed closer to the central star than low excitation, suggesting the existence of
several regions in the inner nebula with different excitation, physical conditions, and/or geometry.

\begin{figure}
\begin{center}
\includegraphics[width=83mm]{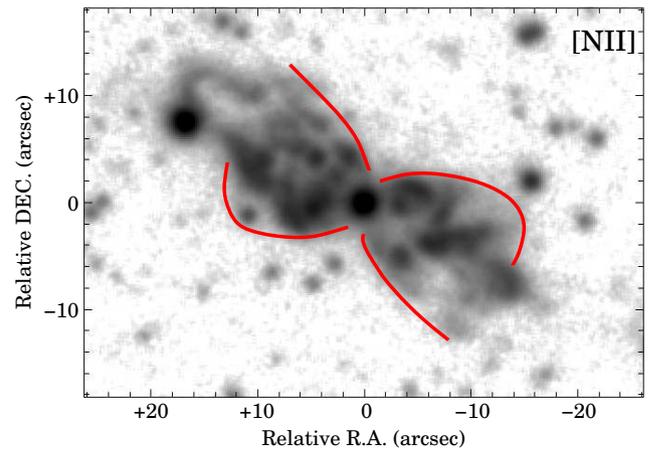}
\caption{Grey-scale reproduction of the [N\,{\sc ii}] NOT image of IRAS18061. The grey levels are logarithmic. The red lines trace the faint emission
  from the bipolar lobes. Otherwise as in Figure\,1.}
\end{center}
\end{figure}

\begin{figure}
\begin{center}
\includegraphics[width=83mm]{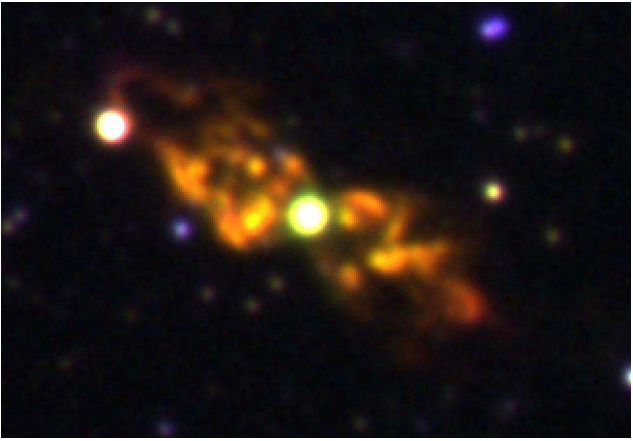}
\caption{Colour composite image of IRAS18061 obtained by combining the
  H$\alpha$ (green), [N\,{\sc ii}] (red), and [O\,{\sc iii}] (blue) images in
 Fig.\,1 (see the text for details). The flux is represented in a linear scale. The size of the 
field shown is 53$\times$37\,arcsec$^2$. Seeing is $\sim$1.1\,arcsec. North is up, east to the left.}
\end{center}
\end{figure}

Figure\,4 shows images of IRAS18061 in different filters and from different epochs: the Palomar Observatory Sky Survey POSS\,I-Blue
and -Red images taken in 1951.6, the POSS\,II-Red image taken in 1996.7, and our CAHA images in the Johnson B, V, and R filters
obtained in 2017.5. The I image is not shown here because it is similar to the rest of the CAHA broad-band images, and the POSS\,II-Blue image
from 1996.7 is not available for this region. The core/inner nebula can be identified in all the images, except in the POSS\,I-Blue one,
while it is faint but clearly present in our B image. This difference cannot be attributed to sensitivity: the POSS\,I-Blue image has a somewhat
better seeing and is deeper than our B image (see Figure\,4). Most probably the internal extinction towards the core was higher in 1951.6 than in later
epochs, hiding the inner nebula at blue wavelengths. We tried to measure the change of the $R-B$ colour between 1951.6 and 2017.5 but the poor
  spatial resolution of the images and the quality of the R image in the POSS\,I did not allow us to obtain a reliable result. Acker et al. (1991)
reported faint H$\beta$ and [O\,{\sc iii}] emission lines in a spectrum obtained in 1989.4. We have inspected this spectrum in the
HASH PN database\footnote{http://202.189.117.101:8999/gpne/} (Parker, Boji\v{c}i\`c \& Frew 2016) but these lines cannot be identified. Nevertheless,
the exposure time of 600\,s for this spectrum is too short for the brightness of IRAS18061. The first unambiguous reported detection of the inner nebula at
blue wavelengths seems to date back to the spectrum obtained in 1994.2 by Su\'arez et al. (2006), that clearly shows the H$\beta$ and
[O\,{\sc iii}]$\lambda$$\lambda$4959,5007 emission lines, among others.

With narrow-band images obtained in three epochs, an investigation of nebular proper motions and variability would seem appropriate. However,
the different characteristics of the filters used in each epoch (Section\,2.1) preclude a proper analysis. We only mention that large flux variability
in the nebula is not recognisable among the three epochs.

\begin{figure*}
\begin{center}
\includegraphics[width=160mm, clip=]{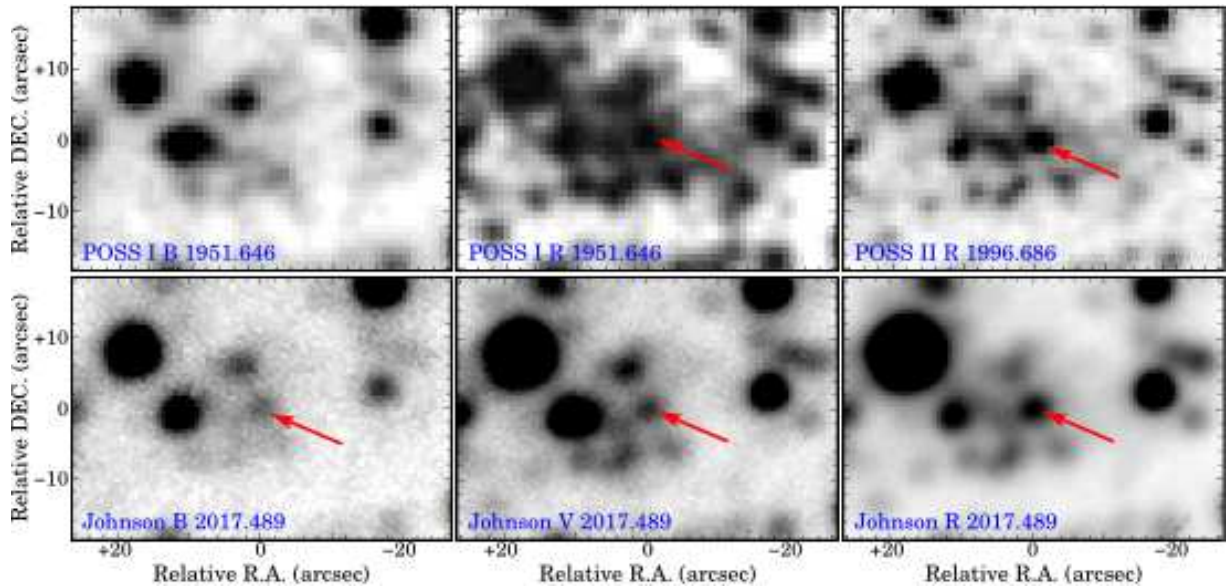}
\caption{Grey-scale images of IRAS18061 (linear flux scale) in different filters and epochs as labelled in 
  the bottom left corner of each panel. The origin (0,0) is at the position of the core that is arrowed, except in the POSS\,I\,B panel where it is
  not detected. Seeing is $\sim$3.5\,arcsec in the 2017 and $\sim$3\,arcsec in the POSS images.}
\end{center}
\end{figure*}

\subsubsection{The bipolar lobes}

The bipolar lobes are observed in the H$\alpha$, [N\,{\sc ii}], and [S\,{\sc ii}] images, but are extremely faint in the [O\,{\sc iii}] one (Figure\,1).
They extend $\sim$24\,arcsec and their main axis is oriented at PA $\sim$+60$^{\circ}$$\pm$1$^{\circ}$. Although the lobes seem to be very narrow in
  Figure\,1, their true extent can be recognised at low intensity levels, as shown in Figure\,2. At PA $\sim$+35$^{\circ}$/215$^{\circ}$ the lobes are disrupted
  and knots and filaments are observed up to $\sim$12$-$15\,arcsec from the centre. Point-symmetry is observed in H$\alpha$ and [N\,{\sc ii}] at PA $\sim$$-$81$^{\circ}$
  and in some bright knots oriented at PA about +60$^{\circ}$. However, bright knots and regions can also be recognized in one lobe, that have
  no counterpart in the opposite one.
  In [S\,{\sc ii}], point-symmetry is difficult to recognized and seems to be restricted to bright knots oriented around PA at $\sim$60$^{\circ}$ and, perhaps, to
  PA $\sim$$-$81$^{\circ}$, and we note that some knots are field stars that are marked in Figure\,1. In general, [N\,{\sc ii}] dominates over H$\alpha$ emission in the lobes,
although variations of their relative intensity are recognisable (Figure\,3). 

\subsubsection{The bow-shock-like structures}

The bow-shock-like structures are clearly seen in the H$\alpha$, [N\,{\sc ii}], and [S\,{\sc ii}] images while only faint emission from
the NE bow-shock can be recognised in the [O\,{\sc iii}] one (Figure\,1). They present noticeable morphological differences
from each other. The NE bow-shock-like structure shows a limb-brightened, bow-shaped morphology extending $\sim$10\,arcsec, that appears displaced
  towards the north with respect to the main bipolar axis. Its tip (hereafter referred to as the NE-BS) is located at $\sim$20\,arcsec from 
the centre and oriented at PA $\sim$+53.2$^{\circ}$$\pm$0.5$^{\circ}$, slightly but clearly different from the main bipolar axis.
The SW bow-shock-like structure (hereafter SW-BS) is more compact, located at $\sim$17\,arcsec from the centre, although faint
emission can be seen up to $\sim$20\,arcsec (Figure\,2), and oriented at PA$\sim$+240$^{\circ}$$\pm$1$^{\circ}$, coincident with the main
bipolar axis. The relative intensity of the [N\,{\sc ii}] emission in the nebula reaches its maximum in the NE- and SW-BS (red colour in Figure\,3).

\subsection{Internal kinematics}

\begin{figure}
\begin{center}
\includegraphics[width=83mm]{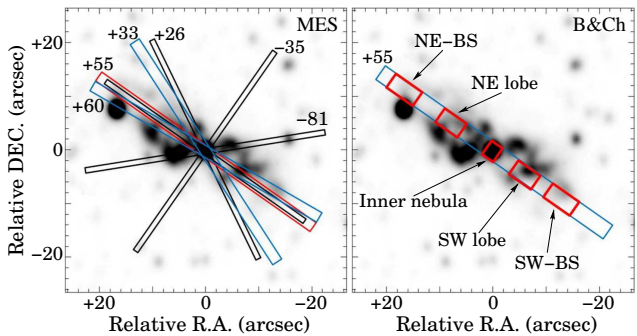} 
\caption{Grey-scale reproduction (linear grey levels) of the Aristarchos [N\,{\sc ii}] image of IRAS18061 superimposed with the slits used for
    long-slit spectroscopy. ({\it left}) The black, red, and blue rectangles correspond to the 2008, 2015, and 2017 slits for the
    MES spectra, respectively. ({\it right}) The blue rectangle corresponds to the slit for the Boller \& Chivens spectra; the red rectangles mark the
    extracted regions for the nebular analysis, that are labelled (Section\,3.3). The PA of the slits is indicated.}
\end{center}
\end{figure}

\begin{figure}
\begin{center}
\includegraphics[width=83mm]{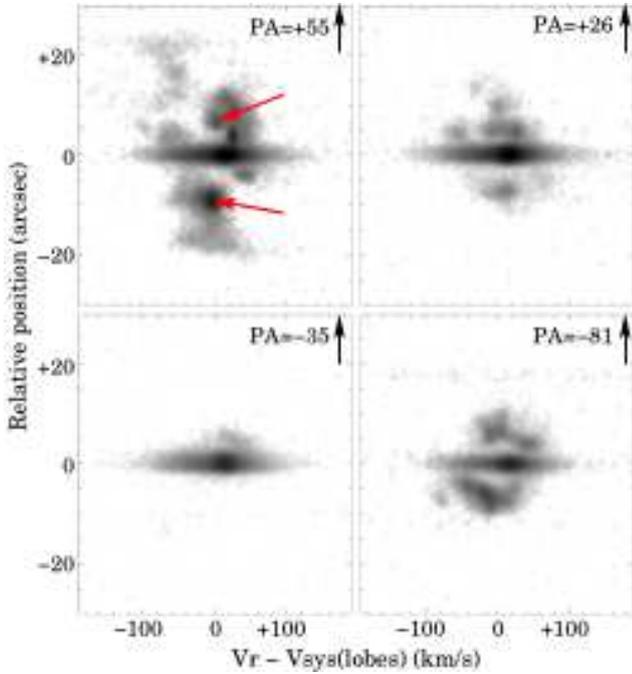} 
\caption{Grey-scale (logarithmic grey levels) PV maps of the [N\,{\sc ii}]$\lambda$6583 emission line from the 2008 MES spectra. A smooth 
with a 3$\times$3\,pixel box has been used for the representation. The slit PA and the orientation are indicated in the upper right corner of each
PV map. The origin of radial velocities is the adopted (LSR) systemic velocity of the bipolar lobes ($V$$_{\rm sys}$(lobes)$\sim$+67\,km\,s$^{-1}$, see 
text). The origin of spatial positions corresponds to the centre of the inner nebula. Two bright ``knots'' discussed in the text are arrowed in the 
PV map at PA +55$^{\circ}$.}
\end{center}
\end{figure}

\begin{figure}
\begin{center}
\includegraphics[width=83mm]{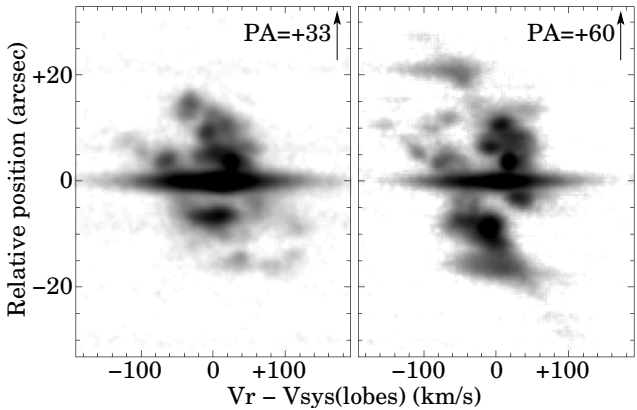} 
\caption{Grey-scale (logarithmic grey levels) PV maps of the [N\,{\sc ii}]$\lambda$6583 emission line from the 2017 MES spectra. Otherwise as
  in Figure\,6.}
\end{center}
\end{figure}

\begin{figure*}
\begin{center}
\includegraphics[width=160mm]{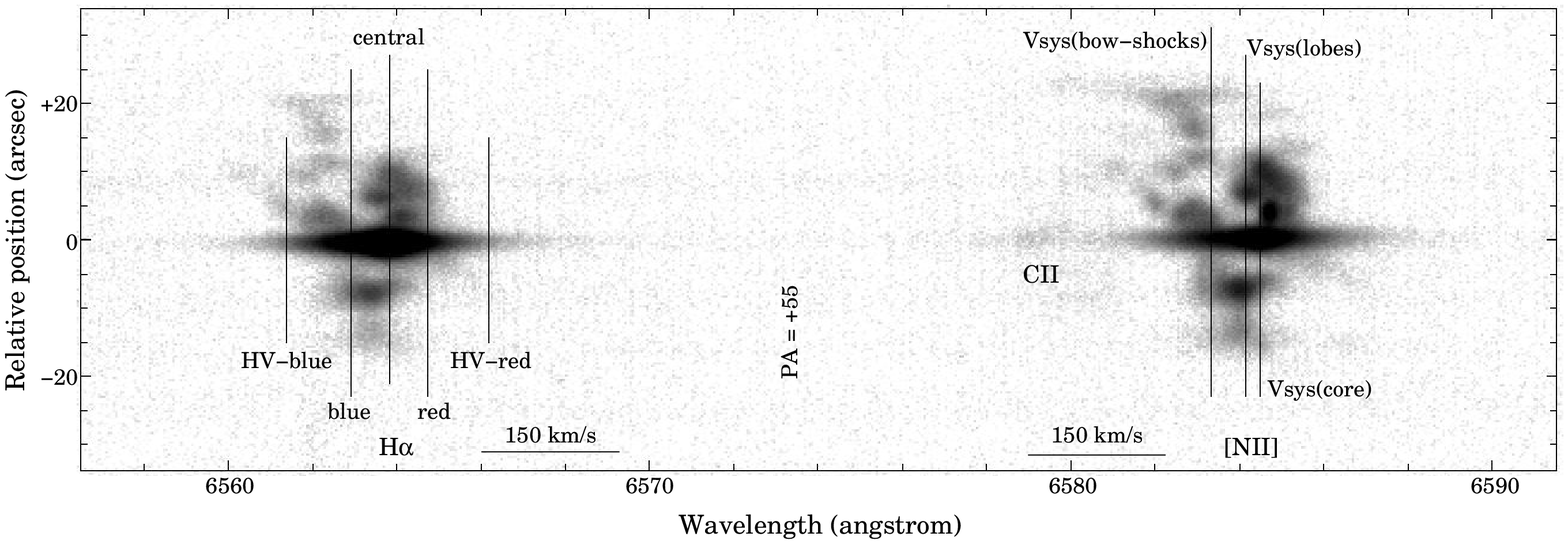}
\caption{Grey-scale (logarithmic grey levels) reproduction of the 2015 MES spectrum at PA +55$^{\circ}$ showing the H$\alpha$ and
  [N\,{\sc ii}]$\lambda$6583 emission lines. The origin of spatial positions corresponds to the centre of the inner nebula. The vertical lines on
  the H$\alpha$ emission line mark the radial velocities of blue, central, red, and HV-components identified in the inner nebula; the vertical lines on the
  [N\,{\sc ii}]$\lambda$6583 emission line mark the systemic/centroid radial velocities of inner nebula, bipolar lobes, and
  bow-shocks (see text). Velocity scales are indicated for the H$\alpha$ and [N\,{\sc ii}]$\lambda$6583 emission lines. The C\,{\sc ii}$\lambda$6578
  emission line is labelled.}
\end{center}
\end{figure*}

Figure\,5\,(left) shows the six long-slits used for the MES spectra, superimposed on the Aristarchos [N\,{\sc ii}] image. Figures\,6 and 7 show
  position-velocity (PV) maps of the [N\,{\sc  ii}]$\lambda$6583 emission line at the PAs observed in 2008 and 2017, respectively, and Figure\,8 presents
  the H$\alpha$ and [N\,{\sc  ii}]$\lambda$6583 emission lines as observed in the PA +55$^{\circ}$/2015 spectrum. The three basic nebular structures
identified in the images have their correspondence in the PV maps, namely, (1) a bright and very broad emission feature at the position of the
inner nebula; (2) extended, knotty emission associated to the bipolar lobes, and (3) emission from the bow-shocks-like structures in the PV maps
at PA +55$^{\circ}$ and +60$^{\circ}$. In the following, we will describe the spatiokinematical properties of the three structures separately.  

\subsubsection{The inner nebula}

\begin{figure*}
\begin{center}
\includegraphics[width=160mm, clip=]{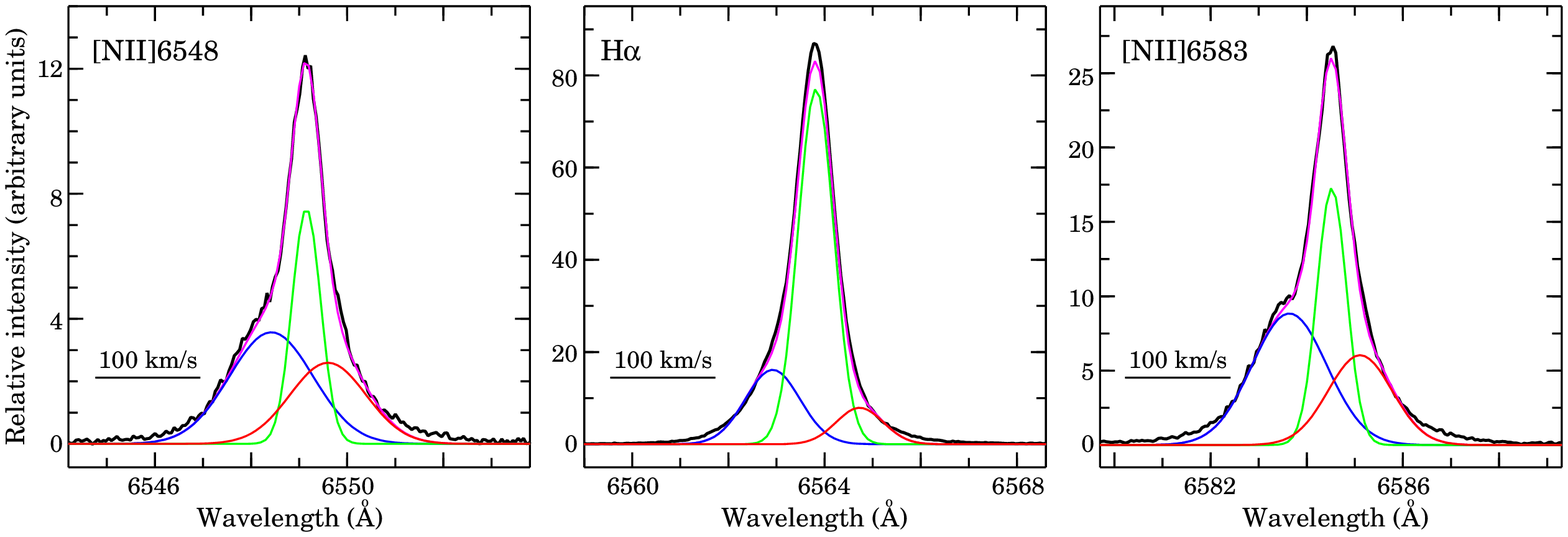}
\caption{[N\,{\sc ii}] and H$\alpha$ emission line profiles of the inner nebula obtained from the long-slit spectrum at 
PA +55$^{\circ}$/2015 (see Figure\,5) and the result of a three-component Gaussian line fit. Black line: observed profile; 
blue, green, and red lines: blue, central, and red components, respectively; fuchsia line: sum of the three fitted 
components (see Table\,1). A velocity scale is drawn for each emission line.}
\end{center}
\end{figure*}

\begin{table*}
\centering
\caption{Properties of the three main kinematical components identified in the inner nebula of $IRAS$\,18061$-$2505.}
\begin{tabular}{lccccccc} 
\hline
Component    &  H$\alpha$   & [N\,{\sc ii}] &  H$\alpha$        & [N\,{\sc
  ii}]        &  H$\alpha$          & [N\,{\sc ii}]  & $I$([N\,{\sc ii}])/$I$(H$\alpha$)\\

             & $V$(LSR)     & $V$(LSR)      & $\Delta$$V$(FWHM) &
             $\Delta$$V$(FWHM)    & Relative Flux$^{(a)}$ & Relative
             Flux$^{(a)}$  &  \\

             & (km\,s$^{-1}$) & (km\,s$^{-1}$)   & (km\,s$^{-1}$) &
             (km\,s$^{-1}$)         &                  &   & \\
\hline

Blue         &   +35$\pm$2   & +48$\pm$3    &   59$\pm$3    & 96$\pm$4      &    0.29$\pm$0.02   & 1.8$\pm$0.3  & 1.6$\pm$0.3 \\

Central      &   +77$\pm$1   & +79$\pm$1    &   40$\pm$2    & 30$\pm$3      &    1.0             & 1.0  & 0.4$\pm$0.1 \\

Red          &   +117$\pm$2  & +102$\pm$5    &   59$\pm$7    & 79$\pm$8    &    0.17$\pm$0.03     & 0.8$\pm$0.2  & 2.2$\pm$0.5 \\

\hline
\end{tabular}

\medskip

$^{(a)}$ Flux(central component)=1.0

\end{table*}

The emission feature due to the inner nebula is not spatially resolved in our spectra, in agreement with its very small
deconvolved angular size (Section\,3.1.1). Figure\,9 shows its [N\,{\sc ii}] and H$\alpha$ emission line
profiles from the PA +55$^{\circ}$/2015 spectrum. Similar profiles are seen in the 2008 and 2017 spectra. The profiles exhibit asymmetries
that suggest the presence of several blended kinematical components, particularly in the [N\,{\sc ii}] emission lines. Satisfactory
fits (but see also below) have been obtained by using a three-component Gaussian line fit with blue, central, and red main components.
The results of the fit are shown in Figure\,9 and Table\,1 lists the values of the (LSR) radial velocity, the velocity width (FWHM), and
the relative flux of the three main components, as averaged from all the H$\alpha$ and [N\,{\sc ii}] emission lines in our spectra.
In Figure\,8 we mark the radial velocity of these components on the H$\alpha$ emission line. The central component presents the same (LSR)
radial velocity in [N\,{\sc ii}] and H$\alpha$ ($\sim$+78\,km\,s$^{-1}$, Table\,1), while the blue and red components
are located symmetrically, at $\sim$$\pm$41 in H$\alpha$ and $\sim$$\pm$27\,km\,s$^{-1}$ in [N\,{\sc ii}], with respect to the central one,
being the velocity separation larger in H$\alpha$ than in [N\,{\sc ii}]. The FWHM of the central component is smaller than those of the blue
and red ones, but larger in H$\alpha$ than in [N\,{\sc ii}]. The blue component is stronger than the red one in both emission lines. The relative
contribution of the components to the total line flux is very different in both emission lines: the central component dominates in H$\alpha$, while
the blue and red components dominate in [N\,{\sc ii}].

Inspection of Figure\,9 reveals that the wings of the emission lines are very extended with a full width at zero intensity (FWZI)
$\sim$400\,km\,s$^{-1}$ that is not accounted for with the three main components, suggesting the presence of additional
very faint high-velocity components (hereafter HV-components). The extended wings are detected in both [N\,{\sc ii}] emission lines in
the 2015 and 2017 spectra but not clearly in the 2008 ones, probably due to their lower S/N ratio (Section\,2.2). In H$\alpha$, the extended wings are
identified in the three set of spectra, probably because the H$\alpha$ emission line from the inner nebula is stronger than
the [N\,{\sc ii}] one (see Section\,3.3.1). We have analysed the residuals of the subtraction between the observed and fitted
profiles but the faintness of the HV-components allowed us to obtain only estimates for some parameters. In particular, the HV-components
seem to be symmetrically located with respect to the central component, with their intensity peaks at $\sim$$\pm$105
and $\sim$$\pm$135\,km\,s$^{-1}$ in H$\alpha$ and [N\,{\sc ii}], respectively, and their flux amounts $\sim$2--4\% that of the central component.
The radial velocity of the HV-components is also marked in Figure\,9 and we emphasize the need of much deeper spectra for a proper analysis. 

To complete our analysis of the emission line profiles, we used the intermediate-resolution spectrum of the inner nebula (Section\,3.3)
and the relative contribution to the total flux of the three main components in each emission line (Table\,1) to obtain the [N\,{\sc ii}]/H$\alpha$ line
intensity ratio in each component. The results are included in Table\,1. The ratio is much higher
in the blue and red components than in central one, suggesting high excitation in the central component. Nevertheless, a high electron
density region exists in the inner nebula (see Section\,3.3.1), in which the [N\,{\sc ii}]$\lambda$$\lambda$6548,6583 emissions 
from the central component may be collisionally de-excited. 

Finally, very faint C\,{\sc ii}\,$\lambda$6578 emission line can be identified in the long-slit spectrum at PA +55$^{\circ}$/2015 (Figure\,6).
A single-component Gaussian line fit to this line gives an (LSR) radial velocity of 80$\pm$3\,km\,s$^{-1}$ and a FWHM of
65$\pm$10\,km\,s$^{-1}$. The radial velocity is similar to that of the central component in [N\,{\sc ii}] and H$\alpha$, but the FWHM
is significantly larger (Table\,1). Given the faintness of the C\,{\sc ii} emission line, we cannot state whether its FWHM is
due to the presence of blue and red components or to a stellar contribution. 

The analysis of the emission line profiles has revealed structures in the inner nebula, that are distinguished by their kinematical
and emission properties. Although our data do not provide information about the relative spatial positions of the 
kinematical components, their symmetry in radial velocity with respect to the central one strongly suggests spatial symmetry, too.
The most simple explanation of the line profiles is that the central component is associated to an inner ring-like structure, the blue and red components
trace inner bipolar lobes, and the HV-components represent a high-velocity bipolar outflow. This interpretation is favoured 
by the high excitation and/or high electron density in the central component (see above), the high [N\,{\sc ii}]/H$\alpha$ intensity ratio
in the blue and red ones, and the faintness of the HV-components. A sketch of the inner nebula is shown in Figure\,A1 where we also mark 
  different regions that are discussed through this paper. With this geometry, the (LSR) systemic velocity of the inner nebula
is $V$$_{\rm sys}$(core)=+78$\pm$1\,km\,s$^{-1}$ that is marked on the [N\,{\sc ii}] emission line of Figure\,8, while 
(projected) expansion velocities of $\sim$20/41/103\,km\,s$^{-1}$ in H$\alpha$ and $\sim$15/27/135\,km\,s$^{-1}$ in [N\,{\sc ii}] are
indicated for the inner ring/inner bipolar lobes/high-velocity outflow. By combining the radial velocities of the inner ring and inner bipolar lobes,
the corresponding deconvolved angular size ($\sim$0.34/0.45\,arcsec in H$\alpha$/[N\,{\sc ii}], Section\,3.1.1), and a distance
of 2\,kpc (Appendix\,B), a kinematical age of $\sim$40--140\,yr is obtained for the inner nebula.
The detection of the inner nebula in the POSS\,II-Red plate from $\sim$1951.6 (Figure\,4) imposes a lower limit of $\sim$70\,yr
for its age, while $\sim$140\,yr may be considered as an approximated upper limit. These numbers imply a very
young structure and we emphasize that the very small angular size and the (projected) velocities in the
inner nebula ensure that its age should accordingly be very small, irrespective of the precise geometry and projection
effects, and for any reasonable distance. 

\subsubsection{The bipolar lobes}

The long-slit at PA $-$35$^{\circ}$ only covers the inner nebula and does not provide information on the bipolar lobes.
At the other PAs, the PV maps are compatible with an expanding bipolar shell such that the NE lobe points  
towards the observer and the SW lobe points away (Figures\,6$-$8). However, there are large differences between the observed PV maps and
those expected from a simple bipolar shell, and, in addition, the NE and SW lobes present kinematical properties very different
from each other.

At PAs +55$^{\circ}$ and +60$^{\circ}$, the NE lobe appears ``closed'' (in the PV maps) by several compact emission features (``knots''),
while the expected redshifted emission from the SW lobe is very faint at PA +55$^{\circ}$ and seems to be absent (or extremely weak)
at PA +60$^{\circ}$. Two point-symmetric knots identified in the images can be recognized at these two PAs and are marked in Figure\,6.
The NE/SW ``knot'' is located at $\sim$+70/+58\,km\,s$^{-1}$ (LSR) and $\sim$7.5/10\,arcsec from the centre. If these two ``knots'' trace a bipolar outflow,
its inclination with respect to the observer is contrary to that of the bipolar lobes. In addition, several ``knots'' are observed in the NE lobe, that
do not have a counterpart in the SW one. 

The PV map at PA +26${^\circ}$ shows five ``knots'' in the NE lobe, including three ``knots'' at about the same spatial position but
different radial velocity, which are incompatible with a simple bipolar shell; the SW lobe presents a more simple kinematics. At PA +33$^{\circ}$,
the SW lobe appears closed whereas the NE lobe appears open in the PV map. These two PAs cover the distorted regions observed in the images
at PAs between $\sim$+25$^{\circ}$ and $\sim$+35$^{\circ}$ and their kinematics is highly suggestive of a bipolar (collimated) outflow that
impacted on and distorted the bipolar lobes.

The PV map at PA $-$81$^{\circ}$ shows point-symmetric features each consisting of two distinct velocity components with maximum radial
velocity splitting close to the inner nebula, that merge at $\sim$7$\arcsec$ from the centre; maximum velocity splitting close to the centre
is not expected in a bipolar shell. Thus, the PV map suggests that a collimated outflow has distorted the the kinematics of the bipolar shell
at PAs around $-$81$^{\circ}$. 

The PV maps clearly show that the emission from the inner nebula, in particular, from its central component, is redshifted with
respect to the apparent radial velocity centroid of the bipolar lobes, suggesting that $V$$_{sys}$(core) is not appropriate as systemic
velocity for the bipolar lobes (Figures\,6--8). We have extracted regions of the bipolar lobes close to the inner nebula
  from the spectra at PAs +55$^{\circ}$, +60$^{\circ}$, +33$^{\circ}$, and $-$81$^{\circ}$ and, from the measured radial velocities, obtained a centroid
  velocity of +67$\pm$3\,km\,s$^{-1}$ (LSR), that will be considered as the systemic velocity of the bipolar lobes $V$$_{sys}$(lobes). The value is similar
to +63$\pm$4\,km\,s$^{-1}$ (LSR) obtained for IRAS18061 from several CO emission lines (Uscanga et al. 2021, in preparation). $V$$_{sys}$(lobes) differs
by $\sim$$-$11\,km\,s$^{-1}$ from $V$$_{sys}$(core) and is marked in Figure\,9.

The analysis of the internal kinematics strengthens the idea that the bipolar lobes are strongly distorted. Their formation seems
to involve several bipolar and non-bipolar collimated outflows at different directions, that impacted the
bipolar shell, resulting in a very complex kinematics. The result of this interaction has been very
different for each lobe. Under these circumstances, it is difficult to predict (or to impose constrains on) how the disrupted
regions move after the interaction process. Moreover, the velocity vector in the ``knots'' could even be non-radial from
the central star. Thus, trying to extract the deprojected velocity and distance to the centre for the ``knots'' is
highly speculative and we will not interpret the PV maps in detail. Nevertheless, it is interesting to obtain the
basic morphokinematical parameters of the bipolar lobes. We have used the tool {\sc shape} to reconstruct the basic structure of the
bipolar lobes (Appendix\,C). Our reconstruction indicates a distorted bipolar shell with main axis at PA +63$^{\circ}$ (close to the PA$\sim$+61$^{\circ}$
deduced from the images), inclination angle respect to the plane of the sky of $i$=15$^{\circ}$, polar radius of 13\,arcsec, and polar
expansion velocity of 163\,km\,s$^{-1}$. Nevertheless, velocities of $\sim$190--200\,km\,s$^{-1}$ are required for the regions along
PA +33$^{\circ}$, which is consistent with acceleration of these regions by the impact of a collimated bipolar outflow. The resulting
kinematical age of the bipolar lobes is $\sim$760\,yr with an estimated error of $\pm$60\,km\,s$^{-1}$. This value is too much larger
than the age of the inner nebula ($\sim$70--140\,yr) to be attributable to uncertainties in the morphokinematic parameters of both
structures, and the difference in ages can be considered real.

\begin{figure}
\begin{center}
\includegraphics[width=80mm, clip=]{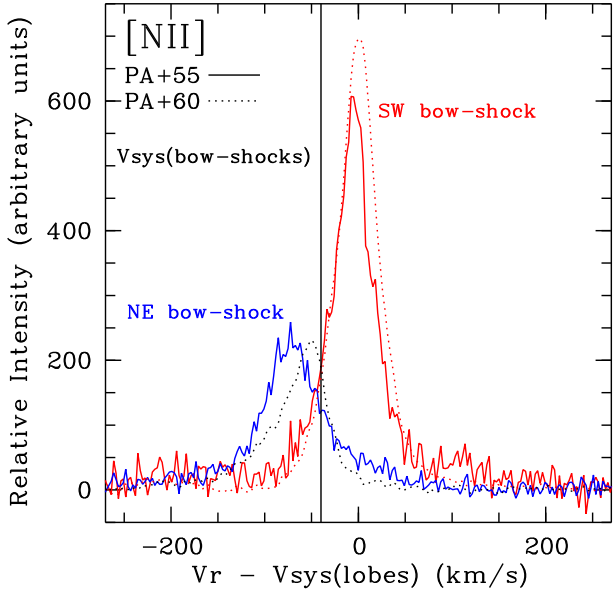}
\caption{Spectral line profiles of the [N\,{\sc ii}]$\lambda$6583 emission
  line at the positions of the NE- and SW-BS as obtained from the long-slit
  spectrum at PA +55$^{\circ}$/2015 by integrating the emission between $-$11 and $-$18\,arsec for the SW-BS and +19 and +22\,arcsec for the
  NE-BS (see Figure\,8). The intensity is arbitrary and the origin of radial velocities is the (LSR) systemic velocity of the bipolar lobes
  $V$$_{\rm sys}$(lobes)=+67\,km\,s$^{-1}$. The vertical line marks the (LSR) systemic velocity of the bow-shocks
  ($V$$_{\rm sys}$(bow-shocks)$\sim$+27\,km\,s$^{-1}$).}
\end{center}
\end{figure}

\subsubsection{The bow-shock structures}

In the PV maps at PA +55$^{\circ}$, the NE-BS appears connected to an elongated ``knot'' and faint redshifted emission, which may be
attributed to the large curved NE bow-shock-like structure (Figure\,1). In the PV map at PA +60$^{\circ}$,
the NE-BS appears somewhat different probably because both slits have covered slightly different regions (Figure\,6). The SW-BS appears
attached to the SW lobe at both PAs +55$^{\circ}$ and +60$^{\circ}$. The NE- and SW-BS present a remarkable triangular shape in the PV maps
with minimum velocity dispersion at their minimum distance to the centre and maximum velocity dispersion approximately
at their maximum distance. Figure\,10 shows their [N\,{\sc ii}] emission line profiles obtained from PAs +55$^{\circ}$/2015 and +60$^{\circ}$. The
FWZI is $\sim$160\,km\,s$^{-1}$ in the SW-BS and $\sim$190\,km\,s$^{-1}$ in the NE-BS. Furthermore, as it can be immediately recognised in
Figures\,5, 6, and 9, both the NE- and SW-BS are blueshifted with respect to $V$$_{\rm sys}$(lobes) and $V$$_{\rm sys}$(core). The intensity
peak of the NE-BS is at $-$9 and +14\,km\,s$^{-1}$ (LSR) in the spectra at PAs +55$^{\circ}$ and +60$^{\circ}$, respectively.
For the SW-BS, we obtain a +55 and +63\,km\,s$^{-1}$ (LSR) in the spectra at PAs +55$^{\circ}$ and +60$^{\circ}$, respectively.
From the values at PAs +55$^{\circ}$ for the NE-BS and PA +60$^{\circ}$ for the SW-BS, that cover better each structure, the (LSR) centroid velocity
of these structures is $\sim$+27\,km\,s$^{-1}$ that is also marked as $V$$_{\rm sys}$(bow-shocks) in Figure\,8 and differs by $\sim$$-$40 and
$\sim$$-$51\,km\,s$^{-1}$ from $V$$_{\rm sys}$(lobes) and $V$$_{\rm  sys}$(core), respectively.  

The properties of the NE- and SW-BS closely resemble those of bow-shocks associated to collimated
jets or bullets in YSOs (e.g., B\"ohm \& Solf 1985). Therefore, the same interpretation probably holds for the NE- and SW-BS. The
FWZI of a bow-shock corresponds approximately to its expansion velocity (Hartigan, Raymond \& Hartmann 1987),
implying that the SW-BS moves at $\sim$160\,km\,s$^{-1}$ and the NE-BS at $\sim$190\,km\,s$^{-1}$. The shape of a
bow-shock on a PV map and its emission line profile depend on the angle it moves with respect to the observer (Raga \& B\"ohm 1986; Hartigan et al. 1987). 
A comparison of the observed profiles with these models shows that the SW-BS moves mainly perpendicular to the line of sight, and the NE-BS moves at $\ga$80$^{\circ}$
with respect to the observer. These results strengthen the existence of noticeable differences between the NE- and SW-BS,
as already noticed in the images, and show that they are not aligned with each other and neither with the main axis
defined by the bipolar lobes. In addition, the large difference between $V$$_{\rm sys}$(bow-shocks) and $V$$_{\rm sys}$(lobes) points out to
particularities in the ejection of the NE- and SW-BS. The kinematical age for these structures is $\sim$950--1000\,yr.

\subsection{Spectral analysis}

\begin{table*}
\centering
\caption{Emission line intensities ($I$(H$\beta$) = 100.0) and derived physical parameters in $IRAS$\,18061$-$2505.}
\begin{tabular}{lcrrrrr} 
\hline
Emission line             & $f$$_{\lambda}$ &  Inner nebula  &  NE lobe          &   SW lobe        &   NE bow-shock   &  SW bow-shock \\ 
\hline
H$\gamma$\,4340           & 0.157        &  43.6$\pm$0.4    &  56.7$\pm$1.6    &   51.6$\pm$2.2   &   ...           &  44.1$\pm$5.9  \\ 

[O\,{\sc iii}]\,4363      & 0.149        &   6.5$\pm$0.3    &   ...            &    ...           &   ...           &   ...           \\ 

He\,{\sc i}\,4471         & 0.115        &   6.1$\pm$0.2    &   ...            &   3.9$\pm$2.0    &   ...           &   ...           \\ 

H$\beta$\,4861            & 0.000        & 100.0$\pm$0.4    &  100.0$\pm$1.2   & 100.0$\pm$1.5    & 100.0$\pm$7.7   & 100.0$\pm$4.1 \\ 

He\,{\sc i}\,4921         & -0.016       &   1.4$\pm$0.1    &   ...            &   ...            &   ...           &   ...         \\ 

[O\,{\sc iii}]\,4959      & -0.026       &  90.3$\pm$0.3    &   10.5$\pm$0.6  &  14.4$\pm$0.8    &   ...           &  54.5$\pm$2.9 \\ 

[O\,{\sc iii}]\,5007      & -0.038       & 274.6$\pm$0.8    &   34.8$\pm$0.7  &  43.0$\pm$0.9    & 380.8$\pm$21.6  & 172.4$\pm$5.7 \\  

[N\,{\sc i}]\,5199        & -0.082       &   1.7$\pm$0.1    &   12.5$\pm$0.6   &  13.7$\pm$0.6    &   ...           &   ...     \\ 

[N\,{\sc ii}]\,5755       & -0.185       &  34.0$\pm$0.1    &    5.6$\pm$0.3  &   5.9$\pm$0.4    &  29.8$\pm$3.9  &  12.2$\pm$1.2 \\ 

He\,{\sc i}\,5876         & -0.203       &  17.7$\pm$0.1    &   13.9$\pm$0.4  &  14.0$\pm$0.4    &   ...            & 14.40$\pm$1.2 \\ 

[O\,{\sc i}]\,6300        & -0.263       &  10.7$\pm$0.1    &   12.1$\pm$0.5  &  19.0$\pm$0.6    &   ...            &   ...      \\ 

[S\,{\sc iii}]\,6312      & -0.264       &  11.8$\pm$0.1    &   ...            &   2.5$\pm$0.3    &   ...            &   ...     \\ 

[O\,{\sc i}]\,6363        & -0.271       &   3.3$\pm$0.1    &    5.2$\pm$0.4  &   3.9$\pm$0.3    &   ...            &   6.8$\pm$1.1 \\ 

[N\,{\sc ii}]\,6548       & -0.295       &  54.5$\pm$0.3    &  196.4$\pm$3.0  & 175.1$\pm$3.2    &  291.9$\pm$27.6  & 239.4$\pm$12.1 \\ 

H$\alpha$\,6563           & -0.298       & 279.7$\pm$1.3    &  292.0$\pm$4.4  & 290.6$\pm$5.3    &  281.8$\pm$26.7  & 286.2$\pm$14.5 \\ 

[N\,{\sc ii}]\,6583       & -0.304       & 168.4$\pm$0.8    &  594.0$\pm$8.9  & 545.5$\pm$10.0   &  954.4$\pm$90.5  & 734.6$\pm$37.3 \\ 

He\,{\sc i}\,6678         & -0.313       &   4.4$\pm$0.1    &    4.7$\pm$0.3  &   3.8$\pm$0.2    &   ...            &   4.7$\pm$0.8 \\ 

[S\,{\sc ii}]\, 6716      & -0.318       &   4.4$\pm$0.1    &   53.5$\pm$0.9  &  45.6$\pm$0.9    &   98.6$\pm$10.0  &  65.9$\pm$3.6 \\ 

[S\,{\sc ii}]\,6731       & -0.320       &   7.2$\pm$0.1    &   68.4$\pm$1.1  &  62.4$\pm$1.2    &  139.3$\pm$14.0  &  81.3$\pm$4.4 \\ 

He\,{\sc i}\,7065         & -0.364       &   7.0$\pm$0.1    &    2.1$\pm$0.2  &   2.6$\pm$0.2    &   ...            &   ...     \\ 

[Ar\,{\sc iii}]\,7136     & -0.374       &  25.4$\pm$0.1    &    4.6$\pm$0.2  &   5.6$\pm$0.2    &   ...            &    12.9$\pm$1.0 \\ 

He\,{\sc i}\,7281         & -0.393       &   0.9$\pm$0.1    &   ...           &   ...            &   ...            &   ...      \\ 

[O\,{\sc ii}]\,7320       & -0.398       &  38.4$\pm$0.2    &   6.6$\pm$0.3   &   7.0$\pm$0.3    &   ...            &   16.4$\pm$1.2 \\ 

[O\,{\sc ii}]\,7330       & -0.400       &  32.5$\pm$0.2    &   2.5$\pm$0.3   &   5.5$\pm$0.2    &   ...            &   14.2$\pm$1.1 \\ 
\hline
$c$(H$\beta$)             &              &  2.54$\pm$0.01   &  1.42$\pm$0.02  &  1.94$\pm$0.02   &  1.43$\pm$0.11   &   1.80$\pm$0.06 \\

log\,$F$(H$\beta$)        &       & $-$13.68$\pm$0.01  & $-$14.37$\pm$0.01  & $-$14.51$\pm$0.01  & $-$15.13$\pm$0.02  & $-$15.00$\pm$0.01 \\ 
(erg\,cm$^{-2}$\,s$^{-1}$) &       &                    &                    &                    &                    &                    \\

$T$$_{\rm e}$[N\,{\sc ii}](K)     &      &   ...            &  8570$\pm$250   &  8970$\pm$300    &  14700$\pm$2700  & 10740$\pm$800    \\    

$T$$_{\rm e}$[O\,{\sc iii}](K)    &              &   16600$\pm$380  &   ...           &   ...            &   ...            &   ...     \\

$N$$_{\rm e}$[S\,{\sc ii}](cm$^{-3}$) &      &   4610$\pm$320   &  1410$\pm$180    &  1820$\pm$280    &   2490$\pm$1460   &  1350$\pm$580 \\

\hline
\end{tabular}
\end{table*}

\begin{table*}
\centering
\caption{Ionic abundances in $IRAS$\,18061$-$2505 relative to H$^+$.}
\begin{tabular}{lccccc} 
\hline
Ion$^a$  & NE lobe           & SW lobe         &  NE bow-shock     &  SW bow-shock         \\ 
\hline

He$^{+}$  &   (10.3$\pm$0.3)$\times$$10^{-2}$    & (9.6$\pm$0.3)$\times$$10^{-2}$    &          ...       &  (10.7$\pm$0.9)$\times$$10^{-2}$ \\ 

O$^{0}$   &   (4.9$\pm$0.6)$\times$$10^{-5}$     & (5.2$\pm$0.7)$\times$$10^{-5}$    &         ...        &  (3.2$\pm$0.2)$\times$$10^{-5}$ \\ 

O$^{+}$   &   (3.0$\pm$0.7)$\times$$10^{-4}$     & (2.5$\pm$0.5)$\times$$10^{-4}$    &        ...         &  (2.3$\pm$0.9)$\times$$10^{-4}$  \\ 

O$^{++}$  &   (2.1$\pm$0.2)$\times$$10^{-5}$     &  (2.2$\pm$0.3)$\times$$10^{-5}$   &  (4.30$\pm$0.23)$\times$$10^{-5}$ &  (4.7$\pm$0.9)$\times$$10^{-5}$ \\ 

N$^{0}$   &   (2.0$\pm$0.3)$\times$$10^{-5}$     &  (2.13$\pm$0.48)$\times$$10^{-5}$ &       ...          &      ...         \\ 

N$^{+}$   &   (1.81$\pm$0.13)$\times$$10^{-4}$   &  (1.44$\pm$0.12)$\times$$10^{-4}$ &  (0.8$\pm$0.3)$\times$$10^{-4}$   &  (1.2$\pm$0.2)$\times$$10^{-4}$  \\ 

S$^{+}$   &   (5.3$\pm$0.4)$\times$$10^{-6}$     &  (4.4$\pm$0.4)$\times$$10^{-6}$   &  (3.5$\pm$0.2)$\times$$10^{-6}$       &  (3.6$\pm$0.7)$\times$$10^{-6}$ \\ 

S$^{++}$  &      ...          &  (0.96$\pm$0.1)$\times$$10^{-5}$   &       ...          &    ...         \\ 

Ar$^{++}$ &   (6.3$\pm$0.8)$\times$$10^{-7}$      &  (6.8$\pm$0.8)$\times$$10^{-7}$    &      ...           &  (10.1$\pm$2.2)$\times$$10^{-7}$   \\ 

\hline
\end{tabular}

$^a$ For ions with more than one transition, an intensity-weighted average has been used.

\end{table*}

\begin{table*}
\centering
\caption{Elemental abundances in $IRAS$\,18061$-$2505 assuming photoionization.}
\begin{tabular}{lccc} 
\hline
Element   & NE lobe          & SW lobe             &  SW bow-shock    \\ 

\hline

He/H      & 0.103$\pm$0.003  &   0.096$\pm$0.003   &  0.107$\pm$0.009  \\ 

O/H       & (3.7$\pm$0.7)$\times$$10^{-4}$      &   (3.23$\pm$0.51)$\times$$10^{-4}$     &  (3.1$\pm$0.9)$\times$$10^{-4}$  \\ 

N/H$^a$   & (2.4$\pm$0.7)$\times$$10^{-4}$      &   (2.1$\pm$0.5)$\times$$10^{-4}$       &  (1.62$\pm$0.83)$\times$$10^{-4}$  \\ 

N/H$^b$   & (2.7$\pm$0.8)$\times$$10^{-4}$      &   (2.3$\pm$0.6)$\times$$10^{-4}$       &  (2.1$\pm$1.1)$\times$$10^{-4}$  \\ 

Ar/H$^a$  & (1.2$\pm$0.3)$\times$$10^{-6}$      &   (1.3$\pm$0.3)$\times$$10^{-6}$       &  (1.9$\pm$0.6)$\times$$10^{-6}$    \\ 

S/H$^a$   & (3.2$\pm$0.2)$\times$$10^{-5}$      &   (1.4$\pm$0.1)$\times$$10^{-5}$       &  (2.2$\pm$0.4)$\times$$10^{-5}$ \\ 
\hline
\end{tabular}

$^a$ICFs from Kingsburgh \& Barlow (1994)

$^b$ICFs from Delgado-Inglada et al. (2014)
\end{table*}

Figure\,5\,(right) shows the long-slit used for the intermediate-resolution spectrum at PA +55$^{\circ}$ that has been used to carry out a spatially
resolved analysis of the emission line intensities, physical conditions and chemical abundances in IRAS18061. We have extracted five nebular
regions from the long-slit spectrum, that are marked in Figure\,5\,right: (1) the inner nebula, corresponding to a region of 2.4\,arcsec in size centred
on the position of its intensity peak; (2) the NE lobe, between 7 and 12\,arcsec from the centre; (3) the NE-BS, between 
17 and 23\,arcsec; (4) the SW lobe, between 5 and 10\,arcsec; and (5) the SW-BS, between 12 and 18\,arcsec. 
The spectrum of the NE-BS is heavily contaminated by the spectrum of a relatively bright field star (Figure\,5) and only some emission lines
could be measured. We note that some faint carbon emission lines due to the [WC8] CS (G\'orny \& Si\'odmiak 2003) are identified in
  our spectrum of the inner nebula, but other lines are not detected. A much deeper spectrum is necessary to analyse the CS. 

The spectra have been analysed with the nebular package {\sc anneb} that uses {\sc iraf}\,2.16, a description
of which can be found in Olgu\'{\i}n et al (2011). Besides, we have checked the results from {\sc anneb} with PyNeb (Luridiana, Morisset \& Shaw 2015). 
Although there are some differences between the two packages (typically $\sim$5--15\% in the ionic abundances, see below), they do not change
the conclusions of the paper. We used the extinction law ($f$$_{\lambda}$) of Cardelli, Clayton \& Mathis (1989) and note that other extinction laws 
do not produce significant changes in the results. For each of the five selected regions, Table\,2 lists the emission line intensities; the logarithmic
extinction coefficient $c$(H$\beta$) obtained from the Balmer emission line ratios assuming recombination case B; the electron density
$N$$_{\rm e}$([S\,{\sc ii}]) derived from the [S\,{\sc ii}]$\lambda$$\lambda$6716,6731 emission lines; and
the electron temperature $T$$_{\rm e}$([N\,{\sc ii}]) or $T$$_{\rm e}$([O\,{\sc iii}]) derived from the
auroral to nebular [N\,{\sc ii}] or [O\,{\sc iii}] emission line ratios R([N\,{\sc ii}]) = 
$I$($\lambda$5755)/$I$($\lambda$6548+$\lambda$6583) or R([O\,{\sc iii}]) = $I$($\lambda$4363)/$I$($\lambda$4959+$\lambda$5007), 
respectively.

\subsubsection{Description of the nebular spectra and physical conditions}

Table\,2 shows that relatively high-excitation emission lines (e.g., [O\,{\sc
  iii}], [Ar\,{\sc iii}]) are stronger in the inner nebula and bow-shocks than in the bipolar lobes. In particular, 
$I$([O\,{\sc iii}])/$I$(H$\beta$) is $\sim$0.45--0.57 in the bipolar lobes, indeed a very
small value. Low-excitation emission lines are weaker in the inner nebula than in bipolar lobes and bow-shocks, reaching in the two
last structures relatively high values of $I$([N\,{\sc ii}])/$I$(H$\alpha$)$\sim$2.5--4.5 and of $I$([S\,{\sc ii}])/$I$(H$\alpha$)$\sim$0.4--0.8.
He\,{\sc ii} and [Ar\,{\sc iv}] emission lines are not detected in our spectra, and the [O\,{\sc iii}]$\lambda$4363 emission line is detected
only in the inner nebula. The spectra indicate a very low-excitation PN and the relatively strong [N\,{\sc ii}] and [S\,{\sc ii}] emission lines
suggest a possible shocks-excitation mechanism.

The extinction ($c$(H$\beta$), Table\,2) is relatively high and reaches its maximum value in the
inner nebula, most probably due to dust and neutral gas that are mainly concentrated
in the core (e.g., Zhang et al. 2012). The extinction is higher in the SW lobe and SW-BS than in their NE counterparts, in
consonance with the inclination of the main nebular axis. 

Electron temperature cannot be obtained for the inner nebula from the [N\,{\sc ii}] emission
lines because R([N\,{\sc ii}])$\sim$0.15 is anomaly high (see also below), although in the bipolar lobes
and bow-shocks $T$$_{\rm e}$([N\,{\sc ii}]) can indeed be obtained and presents lower values in the lobes than in the 
bow-shocks. $T$$_{\rm e}$([O\,{\sc iii}]) can only be obtained for the inner nebula and its value of $\sim$16000\,K 
is typical for PNe. This value differs from that of $\sim$26000\,K obtained by G\'orny et al.
(2004). These authors analysed a single spectrum that probably was a combination of
inner nebula and bipolar lobes spectra. The very different spectra and reddening corrections in these two regions (Table\,2) may 
have led to underestimate, from the combined spectrum, the intensity of the [O\,{\sc iii}]$\lambda$5007 emission line, with respect  
to that of the [O\,{\sc iii}]$\lambda$4363 one only detected in the inner nebula, resulting in a high $T$$_{\rm e}$([O\,{\sc iii}])
and, in consequence, in the peculiar chemical abundances in IRAS18061 (see below).

The electron density $N$$_{\rm e}$([S\,{\sc ii}]) presents moderate values
although it is higher in the inner nebula. Remarkably, evidence for a high electron density region in the inner nebula is
provided by the large value of R([N\,{\sc ii}])$\sim$0.15 that is much higher than those 
usually observed in PNe. Although at low electron densities R([N\,{\sc ii}]) mainly depends on 
$T$$_{\rm e}$, at higher electron densities than the critical 
one for the $^1$D$_2$ level of N$^+$ ($N$$_{\rm c}$$\sim$7$\times$10$^4$\,cm$^{-3}$), R([N\,{\sc ii}]) depends on 
$N$$_{\rm e}$ (e.g., Osterbrock 1974) because collisional de-excitation is more important for
the nebular [N\,{\sc ii}] lines than for the auroral [N\,{\sc ii}] one ($N$$_{\rm c}$$\sim$3.2$\times$10$^7$\,cm$^{-3}$).
Assuming $T$$_{\rm e}$([N\,{\sc ii}]) in the inner nebula in the range found in the other nebular regions ($\sim$9000--16000\,K),
we obtain $N$$_{\rm e}$([N\,{\sc ii}])=5.3--2.0$\times$10$^5$cm$^{-3}$, 
much higher than the value of $N$$_{\rm e}$([S\,{\sc ii}]). This result is corroborated by the radio
continuum data (G\'omez et al. 2008). If the radio continuum emissions at 22 and at 1.665\,GHz arise in
the same region (size $\sim$0.41\,arcsec, Section\,3.1.1), we obtain an electron density of $\sim$2.1$\times$10$^5$\,cm$^{-3}$, 
consistent with the value of $N$$_{\rm e}$([N\,{\sc ii}]). R([O\,{\sc iii}]) is $\sim$0.018, similar to the 
values observed in most PNe, and does not clearly suggest high electron
density. This may be explained because $N$$_{\rm c}$ is $\sim$7$\times$10$^5$\,cm$^{-3}$ for the $^1$D$_2$ level of O$^{++}$
and collisional de-excitation of the nebular [O\,{\sc iii}] lines is not important at the derived $N$$_{\rm e}$([N\,{\sc ii}]),
although some collisional effects cannot be completely discarded. In any case, we will not use
$T$$_{\rm e}$([O\,{\sc iii}]) further.   

The compact core/inner nebula of IRAS18061 resembles that of other pinched-waist PNe as, e.g., Hubble\,12 (Hyung \& Aller 
1996), He\,2-25 and Th\,2-B (Corradi 1995), and K\,4-47 (Gon\c calves et al. 2004), which also present low and high electron
density regions. In all these PNe, the existing spectra do not spatially resolve the different density regions in the core. Therefore, 
the observed emission line fluxes are an unknown combination of fluxes generated in different density regions, with high
electron density playing a crucial role in quenching some emission lines. In these conditions, abundance calculations will give 
unrealistic results, and we will not carry out such calculations for the inner nebula of IRAS18061.

The derived electron densities allow us to obtain a crude estimate for the 
ionised nebular mass. For the inner nebula, we assume a sphere of 0.7\,arcsec in diameter (Section\,3.1.1), and for
electron densities of 4600 and 3.6$\times$10$^5$\,cm$^{-3}$, we obtain an ionised mass between $\sim$2$\times$$\epsilon$$\times$10$^{-5}$ and
$\sim$1.4$\times$$\epsilon$$\times$10$^{-3}$\,M$_{\odot}$, where $\epsilon$ is the filling factor. For each bipolar lobe and
bow-shock, we consider a cylinder of 18.5\,arcsec in length, circular (mean) cross section of 10\,arcsec in size, and electron density of
1600\,cm$^{-3}$ to obtain an ionised mass for the pair of $\sim$5.2$\times$$\epsilon$$\times$10$^{-2}$\,M$_{\odot}$ that may be considered
as an estimate for the total ionised nebular mass.

\subsubsection{Chemical abundances}

Ionic abundances were obtained for the bipolar lobes and bow-shocks, assuming their own values 
of $N$$_{\rm e}$ and $T$$_{\rm e}$ (Table\,2) and are listed in Table\,3. For oxygen, three ionisation 
states are observed and the lack of nebular [Ar\,{\sc iv}] (ionisation potential IP=40.7\,eV) 
makes it highly improbable that O$^{3+}$ (IP=54.93\,eV) exists in the nebula. The abundance of N$^0$
in the main lobes is high, amounting to $\sim$11--15\% that of N$^+$. The few ionic abundances obtained in the
NE-BS may be considered similar to those in the SW-BS, owing to the difficulties to measure the emission lines in
the NE-BS that will not be considered in further calculations. 

Elemental abundances are usually derived from the ionic ones by using ionisation correction factors (ICFs) that
are calculated assuming photoionization. This procedure is not appropriate if shocks contribute to the
nebular excitation (Akras et al. 2020, and references therein), as it could be the case of IRAS18061. Keeping this in
mind, we will obtain elemental abundances by using ICFs, when necessary, and leave for the discussion (Section\,4.1)
the analysis of the excitation mechanism. Elemental abundances are listed in Table\,4 and have been derived as follows.
Because of the lack of nebular [Ar\,{\sc iv}] and He\,{\sc ii} emission lines, helium and oxygen abundances have been
obtained as He/H = He$^+$/H$^+$ and O/H = O$^0$/H$^+$ + O$^+$/H$^+$ + O$^{2+}$/H$^+$, respectively, and we note that no ICF
is needed for oxygen. For nitrogen, we follow both Kingsburgh \& Barlow (1994, hereafter KB94) and
Delgado-Inglada, Morrisset \& Stasi\'nska 
(2014, hereafter D-I+14) to correct for N$^{2+}$ that may exist in the nebula (IP=29.6\,eV), and to the result we add the value of
N$^0$/H$^+$. The nitrogen abundance obtained from D-I+14 is higher by a factor $\sim$1.1 and
$\sim$1.3 in the bipolar lobes and SW-BS, respectively, than that obtained
from KB94. For sulfur and argon we follow KB94 and note that their abundances should be considered with caution because 
they are based in a few emission lines/ionisation states only. In fact, the sulfur abundance in the NE lobe and SW-BS may be  
largely uncertain because the [S\,{\sc iii}]$\lambda$6312 emission line has not been detected in these two structures, which is only
detected in the SW lobe. 

The bipolar lobes and SW-BS present similar helium and oxygen abundances. The argon abundance does not 
present very discrepant values while the sulfur abundance in the SW lobe could be more reliable. Irrespective of the ICF, the
nitrogen abundance is very similar in both lobes and lower in the SW-BS, although the value in the SW-BS has a larger error
and the [N\,{\sc i}]$\lambda$5199 emission line has not been detected. Considering mean values in these structures, the oxygen
abundance is subsolar by a factor $\sim$0.7, and helium and nitrogen abundances are enhanced by a factor $\sim$1.2
and $\sim$3--4, respectively, with respect to solar values (Asplund et al. 2009). The mean nitrogen abundance in the bipolar
lobes (12+log(N/H)$\sim$8.4) and the mean N/O abundance ratio ($\sim$0.7) suggest a type\,I classification for IRAS18061
(Peimbert 1990) and that hot bottom burning (HBB) has taken place in the AGB progenitor. The mean helium abundance is $\sim$0.102,
lower than but still within the range found in other type\,I PNe. A comparison with other PNe shows that IRAS18061 does not
present peculiar abundances (KB94).

\section{Discussion}

\subsection{The progenitor star of $IRAS$\,18061--2505}

As already mentioned, H$_2$O-PNe have been associated to intermediate-mass MS progenitors. A lower limit
of $\sim$4\,M$_{\odot}$ is usually assigned to the initial mass, following the models by, e.g., Karakas \&
Lugaro (2016), and Marigo et al. (2017) that predict a lower limit of $\sim$4\,M$_{\odot}$ for HBB. 
Nevertheless, models by, e.g., Miller Bertolami (2016) predict a lower limit of $\sim$3\,M$_{\odot}$. 
Recent observations of the PN M31\,B477-1 by Davis et al. (2019) provide evidence that HBB has occurred in the AGB phase of
its $\sim$3.4\,M$_{\odot}$ MS progenitor. Intermediate-mass MS progenitors are also suggested by the low Galactic 
latitudes ($|b|$$\la$$2\rlap.^{\circ}6$) and high extinction at visible wavelengths of these objects, although the amount of 
extinction is quite different among H$_2$O-PNe. For instance, IRAS18061 is completely visible (now, but not in $\sim$1951.6, Figure\,4),
K\,3-35 is highly extincted towards its equatorial region but its bipolar lobes are clearly visible (e.g., Blanco et al. 2014),
and $IRAS$\,15103$-$5754 is not optically visible. A possible shortcoming for an intermediate-mass MS progenitor in IRAS18061 is
the small ionised nebular mass. Near-, mid- and far-IR data, and the water masers suggest that noticeable amounts of dust
and molecular gas should exist in the object, most probably in the core, although their masses have yet to be estimated (see below).
The nitrogen abundance and N/O ratio in IRAS18061 are compatible with HBB, if shocks do not contribute to the nebular excitation.

Alternatively, oxygen-rich AGB stars may have low-mass MS progenitors ($M$$\la$1.25\,M$_{\odot}$), a possibility that would be favoured for IRAS188061
by its small ionised mass. However, current evolutionary models (but see below) are not compatible with a low-mass MS progenitor. A $\la$1.25\,M$_{\odot}$
MS progenitor needs $>$7000\,yr (choosing the fastest evolution, Miller Bertolami 2016, hereafter
MB16; see also Vassiliadis \& Wood 1994; Bl\"ocker 1995) to evolve from the AGB to the PN phase, and the resulting PN
could be expected to present characteristics of an evolved PN. The electron densities in IRAS18061 are not typical of evolved PNe (Miranda et al. 2017 and
references therein), and its small kinematical age ($\sim$1100\,yr) is not very suggestive of an evolved PN. Moreover, if H$_2$O-PNe were related to
  low-mass MS progenitors, one could expect to detect more H$_2$O-PNe than the very small number of identified objects. Finally, if shocks do not
contribute to the nebular excitation, the chemical abundances are incompatible with a low-mass MS progenitor. 

An estimate for the MS progenitor mass of a CS may be obtained by comparing the nebular 
abundances with models for stellar yields. However, this comparison requires that photoionization dominates the
nebular excitation and, hence, that the use of ICFs is appropriated to obtain the elemental abundances. The
strong low-excitation emission lines observed in some PNe (e.g., Gon\c calves et al. 2009), including IRAS18061, have been explained by shocks
in a highly ionised medium (Dopita 1997). Nevertheless, photoionization models are also able to reproduce the observed strong low-excitation emission
lines, as recently shown by Akras et al. (2020). We have compared several line intensity
ratios in the bipolar lobes of IRAS18061 with the diagnostic diagrams by Akras et al. (2020, their Figures\,5 to 7) and they are well
reproduced with their photoionization models, while the line intensity ratios in the SW-BS seem to be more compatible with shock-excitation. These authors
assume a CS with $T$$_{\rm  eff}$=1--2$\times$10$^5$\,K to compute their models. The $T$$_{\rm  eff}$ of the CS of IRAS18061 is unknown but most probably
$<$6$\times$10$^4$\,K that would be more favourable to produce strong low-excitation emission lines than models with higher $T$$_{\rm  eff}$. 

\begin{figure}
\begin{center}
\includegraphics[width=80mm, clip=]{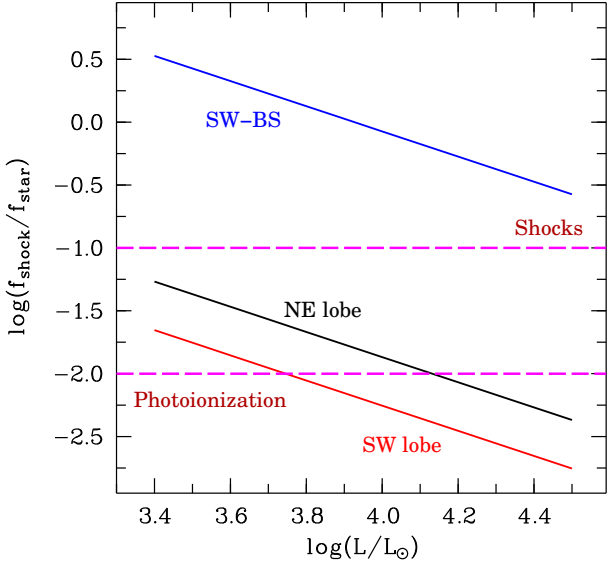}
\caption{Diagram of the ratio of flux due to shocks (f$_{shocks}$) to flux due to
  the central star (f$_{star}$) as a function of the stellar luminosity for
  the NE and SW lobes, and SW-BS. The dashed horizontal 
lines mark the shock-excited and photoionised regimes that 
are separated by a region where both shocks and photoionization contributed 
to the emission (see text for details). } 
\end{center}
\end{figure}

To check further the possible contribution of shocks in IRAS18061, we will follow the
prescription by Akras \& Gon\c calves (2016) who use the photon flux relation (or the ratio of fluxes, Lago et al. 2019) due to
shocks ($f$$_{\rm  shock}$) and central star ($f$$_{\rm star}$) to distinguish the excitation mechanisms. These
authors found that for values of log($f$$_{\rm shock}$/$f$$_{\rm star}$)$>$$-$1 the nebular excitation is
dominated by shocks, for values $<$$-$2 photoionization dominates, whereas for intermediate values both shocks and photoionization
contribute to the excitation. For IRAS18061 we will obtain log($f$$_{\rm shock}$/$f$$_{\rm star}$) as a function of the stellar luminosity
for the bipolar lobes and SW-BS (Lago et al. 2019, their equation\,3), considering luminosities 
log({\it L}/L$_{\odot}$) between 3.4 and 4.5 that approximately cover the whole mass range of 
central stars (VW94; B95; MB16), and in each structure its distance to the centre, electron density, and expansion velocity that is
assimilated to the shock velocity. In the case of the NE and SW lobes, we will consider the middle point where their
spectra have been extracted as distance to the central star (Section\,3.3), and as shock velocity at that distance, the velocity
deduced from the spatiokinematical model: 9.5\,arcsec/55\,km\,s$^{-1}$ for the NE lobe, and 7.5\,arcsec/44\,km\,s$^{-1}$ for the SW lobe.

The resulting log($f$$_{\rm shock}$/$f$$_{\rm star}$) vs. log({\it L}/L$_{\odot}$) diagram is shown in
Figure\,10. The lines corresponding to the NE and SW bipolar lobes appear displaced from each other in Figure\,10
due to the different distance to the centre and shock velocity considered for each lobe. 
Shock excitation appears as the dominant excitation mechanism for the SW-BS (and NE-BS, not shown here) for the whole range of
luminosities, in agreement with their bow-shock nature. For log({\it L}/L$_{\odot}$)$\la$3.75, corresponding to a
$\la$1.5\,M$_{\odot}$ MS progenitor (although the precise value of the mass depends on the model, see MB16, his Figure\,11),
the emission from the bipolar lobes contains contributions from shock-excitation and photoionization. For log({\it L}/L$_{\odot}$)$\ga$4.15,
corresponding to a $\ga$3\,M$_{\odot}$ MS progenitor, the emission from both lobes is dominated by photoionization.
According to these results, photoionization alone may explain the spectra of the bipolar lobes if the initial mass is $\ga$3\,M$_{\odot}$.
Let us check the results suggested from the chemical abundances, assuming that the use of ICFs has been appropriated. We have
compared the helium and nitrogen abundances, and N/O abundance ratio with several models for stellar yields (e.g., Marigo 2001;
Karakas 2010; Karakas \& Lugaro 2016; MB16), assuming the oxygen abundance as representative of the
initial metallicity. Although the models are very different from each other, and use different hypothesis and treatments,
the comparison suggests MS progenitor masses between $\sim$3 and $\la$4.2\,M$_{\odot}$, excludes masses $\ga$4.5\,M$_{\odot}$, and points to a
subsolar initial metallicity $\sim$0.008-0.01. The mass range is approximately bracketed by the lower limits for HBB in the different models 
and is compatible with the luminosities required for photoionization (Figure\,10). It is worth noting that most models predict a drop in the
helium abundance around 3--4\,M$_{\odot}$, compatible with the apparent low helium abundance in IRAS18061. We also note that the minimnum 
MS progenitor mass required for photoionization derived from Figure\,10 and that obtained from the chemical abundances are based in two
methods that are independent from each other.

The diagram in Figure\,10 has been obtained assuming a distance of 2\,kpc for IRAS18061 but it is interesting to discuss
the results for other distances. For values $<$2\,kpc, the lines in Figure\,10 shift downwards and the minimum
mass required for photoionization decreases. However, masses between $\sim$1.25 and $\la$3\,M$_{\odot}$ are incompatible with an oxygen-rich
PN, as it is indicated by the water masers in IRAS18061. For values $>$2\,kpc, the lines in Figure\,10 shift upwards and the
minimum mass increases. These masses lead to higher He and N abundances, and N/O ratio than the values obtained in IRAS18061
(e.g., Karakas \& Lugaro 2016), resulting incompatible with the abundances in IRAS18061. 

The results discussed above show that a $\sim$3--4\,M$_{\odot}$ MS progenitor is able to account for the chemical abundances in
IRAS18061, which is in agreement with the expectation of intermediate-mass MS stars as progenitors for H$_2$O-PNe (but see below). If so, as already
mentioned, large amounts of dust and neutral material should exist in IRAS18061. In this respect, it is worth 
noting ``water fountains'' (WFs), post-AGB stars with high velocity water masers tracing very young jets (e.g., Imai 2007),  that have
been suggested to be the immediate precursors of H$_2$O-PNe, which is strongly supported by {\it IRAS}\,15013---5754 that presents both WF
and PN characteristics (Su\'arez et al. 2015; G\'omez et al. 2015a, 2018b). Alike H$_2$O-PNe, WFs have also been associated with
$\sim$4--8\,M$_{\odot}$ progenitors (e.g., Su\'arez et al. 2008; Young et al. 2011; Imai et al. 2012; Rizzo et al. 2013). Models for 
the spectral energy distribution of WFs seem to require the existence of massive dusty toroids/rings ($\sim$0.5--2.5\,M$_{\odot}$) in these objects
(Dur\'an-Rojas et al. 2014). An analysis of the SED of IRAS18061 would allow us to estimate the dust and gas mass. However, making 
  a realistic SED model for IRAS18061 is far from simple. The stellar spectrum, $T$$_{\rm eff}$, and the precise density distribution in
  the inner nebula are unknown; amorphous carbon, graphite, and silicates with different compositions and grain sizes should be included and tested in
  the model. In consequence, a dedicated study should be carried out to obtain reliable values for the dust and gas masses, which is beyond the scope of this paper.
Finally, the results from Figure\,10 impose some constrains to the distance of the object, that should be around 2\,kpc to make compatible the observed
chemical abundances, the MS progenitor mass, and the oxygen-rich nature of the object.

Even though the results obtained above are consistent with a $\sim$3--4\,M$_{\odot}$ MS progenitor, it should be emphasized that the role of shocks in
  the nebular excitation of PNe is a very complex problem that is not solved yet. Therefore, a low-mass MS progenitor for IRAS18061 cannot
  be conclusively ruled out, and we will consider this possibility at some points of the discussion below.

\subsection{An oxygen-rich neutral ring in the core of IRAS18061}

Water maser emission is always observed towards the central regions of H$_2$O-PNe, although in K\,3-35 and {\it IRAS}15103-5754
it has also been detected associated to jets (Miranda et al. 2001a; Su\'arez et al. 2015; G\'omez et al. 2015a, 2018a). Moreover,
water masers in H$_2$O-PNe tend to be distributed mainly perpendicular to the main nebular axis, suggesting that they are associated to
equatorial rings, as in K\,3-35 (Uscanga et al. 2008), {\it IRAS}\,17347$-$3139 (Tafoya et al. 2009), and {\it IRAS}\,15103$-$5754
(G\'omez et al. 2018a), and was also proposed for IRAS18061 (G\'omez et al. 2012). The water masers in IRAS18061 are observed at $\sim$0.02--0.05\,arcsec
towards the southwest from the CS, oriented approximately at PA$\sim$160$^{\circ}$, and with (LSR) radial
velocities between $\sim$+57 and $\sim$+64\,km\,s$^{-1}$ (G\'omez et al. 2008). These properties, when combined with the
morphokinematic ones of the bipolar lobes, provide solid arguments for the existence of a neutral ring in the core of IRAS18061
traced by the water masers. The water masers are blueshifted with 
respect to $V$$_{\rm sys}$(lobes) [and $V$$_{\rm sys}$(core)], distributed approximately perpendicular to the major nebular axis,
and projected onto the (redshifted) SW lobe, as expected if they arise in the front part of a ring that is tilted in the same
sense as the main bipolar axis. Furthermore, a ring-like structure and not a spherical distribution of neutral material
in the core, is required to detect the CS (G\'orny \&  Si\'odmiak 2003). The oxygen-rich ring is drawn in Figure\,A1 and
we note that it would be compatible with the red component of the inner nebula being fainter than the blue one.
From the positions of the water masers in G\'omez et al. (2008, 2012), a crude estimate of the ring radius is $\sim$0.1\,arcsec, corresponding
to $\sim$200\,AU, similar to the radius of water maser rings in other H$_2$O-PNe. Water maser emission requires high temperatures of $\sim$100-400\,K
and high particle densities of $\sim$10$^{6-9}$\,cm$^{-3}$ (Hollenbach, Elitzur \& McKee 2013). Therefore, the derived radius probably corresponds
to the innermost hottest and densest region of the neutral ring (Figure\,A1), although the ring may be larger. 

The water masers undoubtedly indicate that the neutral ring is oxygen-rich. This agrees with the scenario proposed by Cohen et
al. (1999) that oxygen-rich material in PNe with late-type [WC] central stars resides in a ring which could also be the place where the
crystalline silicate emission from IRAS18061 arises (Perea-Calder\'on et al. 2009). This ring should have been formed during 
the oxygen-rich phase of the AGB progenitor of IRAS18061, before the transition to a carbon-rich chemistry. According to Cohen et al. (1999),
  carbon-rich material should be interior to the oxygen-rich ring and the inner nebula may be expected to be carbon-rich.

\subsection{The formation of $IRAS$\,18061--2505}

Binary/multiple central stars provide a scenario in which many properties of PNe may be accounted for, at least qualitatively (see, Boffin \& Jones
2019 for a recent review). Particularly relevant is the case of binary CSs that evolved in a common envelope (CE). At
least $\sim$20\% of PNe may be the ejected CE (Miszalski et al. 2009) and bipolarity has been suggested as a possible signature of
common envelope evolution (CEE, Reichardt et al. 2019). Numerical simulations of CEE are able to reproduce bipolar PNe (e.g., Garc\'{\i}a-Segura,
Ricker \& Taam 2018; Reichardt et al. 2019 and references therein). The bipolar lobes and water maser ring in IRAS18061 are compatible with
the characteristics expected from CEE. The high particle density in the oxygen-rich ring, as compared with the low density in the bipolar lobes,
strongly suggests a very high density contrast in the AGB envelope, with most material concentrated at the equatorial/orbital plane, and very low
density regions above and below that plane. When the fast wind from the CS interacts with such a highly anisotropic density distribution, it will
be strongly decelerated in the equatorial plane while it will encounter less resistance along the polar regions, resulting in a pinched-waist PN,
in agreement with the morphology of IRAS18061. A very high density contrast suggests a relatively low mass ratio between
CS and companion (e.g., Zou et al. 2020) and that the presumable companion is not a very low-mass star. Moreover, bipolar and asymmetric jets from the 
secondary may also be ejected during CEE at different directions, shaping the CE (Soker 2019; Frank et al. 2018;
Shiber et al. 2019) and contributing to the large differences between the two bipolar lobes and their strongly
disrupted morphology. Nevertheless, collimated outflows could be ejected after CEE, contributing to further disruption of the bipolar lobes.

The properties of the NE-, SW-BS do not suggest that they have played a role in the formation of the
bipolar lobes. The NE- and SW-BS are the oldest component identified in IRAS18061 and their formation may have preceded that of the bipolar lobes 
and, therefore, occurred before CEE. Although several scenarios are possible (Blackman \& Lucchini 2014), an interesting possibility for IRAS18061  
is that the NE- and SW-BS may be related to Roche-lobe overflow or grazing
envelope evolution that may result in the formation of an accretion disk around the companion, from which
collimated outflows are ejected (Soker 2015;  Shiber, Kashi \& Soker 2017; Shiber et al. 2019). Precession/wobbling and/or non strictly simultaneous
ejection from each side of the accretion disk (Vel\'azquez et al. 2014), and/or contribution of the orbital velocity to the velocity of the
collimated ejecta (Miranda et al. 2001b,c) could account for the morphological differences between the NE- and SW-BS, their non-relationship to
the axis of the bipolar lobes, and the difference between $V$$_{\rm sys}$(bow-shocks) and $V$$_{\rm sys}$(lobes). 

The formation of the inner nebula may be understood in the context of the [WC] nature of the CS, which strongly 
suggests a thermal pulse (e.g., Bl\"ocker 2001). The large amounts of material ejected in a thermal pulse form a new
PN or shell which, in IRAS18061, can be identified with its inner nebula. The youth of IRAS18061 makes it attractive the idea
of a final thermal pulse at the very end of the AGB. However, the IRAS fluxes of IRAS18061 are incompatible with those of IR-[WC] stars
that are expected to result from a final thermal pulse (Zijlstra 2001). In particular, the flux in the IRAS bands of IRAS18061 is
  $\sim$0.36$\times$10$^{-11}$\,W\,m$^{-2}$ (Iyengar 1987), lower than 0.8$\times$10$^{-11}$\,W\,m$^{-2}$ required for IR-[WC] stars, and its
  position in the logF(60)/F(25) vs. log(F(25)/F(12) and logF(60)/F(12) vs. [WC] subclass diagrams does not coincide with that of IR-[WC]
  stars, as defined by Zijlstra (2001). A late thermal pulse (LTP) on the post-AGB horizontal track 
or a very late thermal pulse (VLTP) on the cooling track are then the options, in both cases suggesting a born-again scenario
(e.g., Bl\"ocker 2001; Herwig 2001; Miller Bertolami et al. 2006; see also below).

A born-again scenario is favoured by several results. After an LTP/VLTP, the CS returns to the AGB in a few hundreds to a few years
(e.g., Bl\"ocker 1995; Hajduk et al. 2005; Miller Bertolami et al. 2006), and large amounts of dust are formed, hiding the CS at optical
wavelengths until reheating of the central star and/or shocks destroy the dust and the CS is seen again (Seitter 1987; Kerber et al. 2002; Hajduk et al. 2005;
Hinkle et al. 2008; Rechy-Garc\'{\i}a et al. 2020). The non-detection of the inner nebula in $\sim$1951.6 (Figure\,4) suggests large amounts of dust in the
core and that the CS revisited the AGB, placing the occurrence of the thermal pulse sometime before $\sim$1951.6. The decrease of the extinction
from $\sim$1951.6 to $\sim$1994.2 indicates that dust has been destroyed during those years and that the CS is
reheating. The high velocity material ejected in a thermal pulse causes shocks that propagate in the surrounding material  (Guerrero et al. 2018).
Evidence for these shocks is provided by the expansion velocities in the inner nebula that are higher in H$\alpha$ than in [N\,{\sc ii}], while
the opposite is expected in a photoionised nebula. Shocks may also cause an inverted ionisation structure (Guerrero et al. 2018) that is
partially recognised in the inner nebula, where the [O\,{\sc iii}]
emission shows a larger (deconvolved) size than the [S\,{\sc ii}] emission. The H$\alpha$ and [N\,{\sc ii}] emissions do not follow this behaviour, although the
ring and bipolar lobes of the inner nebula contribute very differently to the H$\alpha$ and [N\,{\sc ii}] emission lines, and their 
(deconvolved) size could be dominated by different structures. Finally, shocks (and photoionisation) may contribute to the ionization of the innermost
layer of the oxygen-rich ring, resulting in a high electron density region which may be associated with the inner ring/central component (see Figure\,A1). 

A possibility to distinguish between LTP and VLTP in IRAS18061 is comparing the evolutionary time scales of its CS and the age of its bipolar
  lobes. However, if CEE has occurred in IRAS18061 as proposed above, evolutionary models constructed for single CSs cannot be appiled to its CS (Miller Bertolami
  2019). In particular, a CE may be ejected in a several years or a few decades (Chamandy et al. 2020), drastically reducing the transition times\footnote{the duration of
  the early AGB phase, from the end of AGB to $T$$_{\rm eff}$$\sim$10000\,K (MB16).} that single star models predict to be $\sim$1400--930\,yr for a 3--4\,M$_{\odot}$ MS
  progenitor (MB16). If we assume that, after (or a short time after) the ejection of the CE, the CS evolves approximately as predicted by single star models,
  the crossing times\footnote{the duration of the late AGB phase, from $T$$_{\rm eff}$$\sim$10000\,K to the maximum $T$$_{\rm eff}$ attainable on the horizontal track (MB16).}
  for a 3--4\,M$_{\odot}$ MS progenitor are $\sim$340--70\,yr (MB16), much smaller than the age of the bipolar lobes, placing the CS on the cooling track at
  the instant of the thermal pulse, and favouring a VLTP. For a low-mass MS progenitor of $<$1.5\,M$_{\odot}$, the crossing time is $>$3000\,yr,
  larger than the age of the bipolar lobes, favouring an LTP. In any case, a definitive distinction
between LTP and VLTP in IRAS18061 requires obtaining the chemical abundances in the CS, that are determined by the instant when the thermal pulse occurs (Herwig 2001). 

High velocity outflows and rings are systematically observed in the ejecta of born-again CSs (Zijlstra 2002 and
  references therein; Borkowski et al. 1993; Hajduk et al. 2005; Hinkle et al. 2008; Chesneau et al. 2009; Hinkle \& Joyce 2014; Fang et al. 2014;
  Rechy-Garc\'{\i}a et al. 2020). Small bipolar nebulae have also been reported in V\,605\,Aql and
V\,4334\,Sgr (Sakurai's object) (Hahduk et al. 2005; Hinkle et al. 2008; Hinkle \& Joyce 2014. These structural components are also identified in the inner
nebula of IRAS18061, suggesting a common formation mechanism for all born-again ejecta. Peculiar in this PN is the pinched-waist bipolar
  morphology of the (``old'') main shell that, to the best of our knowledge, is not observed in the ``old'' shells of other PNe with born-again CSs. Remarkably,
  the main shell of IRAS18061 also presents a ring,
  bipolar lobes, and high-velocity outflows, as the inner shell. As discussed above, CEE is a plausible interpretation for the formation of
the oxygen-rich ring and bipolar lobes, and has also been suggested to explain the small disk around Sakurai's object, and, perhaps, the radial velocity
variations in the born-again CS of Abell\,30 (Chesneau et al. 2009; Jacoby et al. 2020). Moreover, several scenarios invoke a companion to explain the
dual chemistry of [WC] CSs (De Marco \& Soker 2002; De Marco 2008). From these results, we
speculate that IRAS18061 could be a case of double CEE, in which the formation of the inner nebula would be related to a second CEE that occurred when the
CS expanded in its return to the AGB, and engulfed the presumable close companion that may have resulted from the first CEE originating
the main shell. It is worth to mention that the systemic velocities of the bipolar lobes and inner nebula differ by $\sim$11\,km\,s$^{-1}$. This may be due
to disruption of a companion, asymmetric ejection of a CE, ejection of a third body in the system, or a kick on the binary caused by the ejection of jets
(Soker 2016 and references therein; Shiber et al. 2009). Although we cannot favour any of these possibilities in IRAS18061, all of them involve a
companion (or companions) to the CS, which is also supported by other properties of this PN. Therefore, differences in the systemic/centroid
velocity of different structures in a PN could be indicating the possible presence of a binary/multiple CS, as already suggested by Miranda et al.
(2001b; see also Miranda et al. 2001c; G\'omez et al. 2018b).

The kinematical age of the bipolar lobes ($\sim$760\,yr) is much larger than the survival time of $\sim$100\,yr for water
maser emission beyond the AGB (Lewis 1989; G\'omez et al. 1990), and remnants of the AGB water masers should not exist in IRAS18061. Although the value
of $\sim$100\,yr should be considered as an approximation, values larger than a few hundred years should not be expected because water masers are not
typically found in PNe (de Gregorio-Monsalvo et al. 2004; G\'omez et al. 2015a). The water masers in IRAS18061 may be associated with the recent formation
of the inner nebula. We propose that water maser emission in IRAS18061 has been reactivated late in the post-AGB evolution through shocks in the pre-existing
oxygen-rich ring, that are generated by the thermal pulse. This scenario would be favoured if the central star evolves fast
enough before the LTP/VLTP, such that the oxygen-rich ring has not expanded very much and still maintains the high particle density
required to pump water masers when the thermal pulse occurs. This requirement may easily be fulfilled by an
intermediate-mass MS progenitor, as the case of IRAS18061 seems to be, while in the case of a low-mass MS progenitor, it would depend
on the evolutionary time scale. If an LPT/VLTP lately reactivates water maser emission,
the resulting H$_2$O-PN will not be at the very moment of its first entrance in the PN phase and IRAS18061 cannot be considered
  as a very young PN, although its inner nebula indeed is extremely young. In consequence, the presence of water maser emission may not
necessarily indicate always an extremely young PN.

\subsection{Implications for the formation and evolution of H$_2$O-PNe and related objects}

IRAS18061 gathers many characteristics that pertain to a variety of phenomena and structures observed in PNe, including
narrow waist bipolar lobes, collimated outflows, multiple ejections, rings, and could be a key object to understand the
formation of complex PNe. In what follows, we will concentrate on the possible implications of our results to understand
other H$_2$O-PNe and possibly related objects.

As already mentioned, jet-envelope interaction may excite water masers in very young PNe. On the other hand, it has been suggested
  that water masers in PNe are the remnants of the AGB ones. According to single star evolutionary models (B95; MB16), MS progenitors
  with $\ga$6\,\,M$_{\odot}$ should be required, that would already be on the cooling track, owing their extremely fast evolution.
  However, the properties of H$_2$O-PNe strongly suggest that they host binary/multiple CSs. If the formation of H$_2$O-PNe is associated with CEE, as
  suggested by IRAS18061 (see also G\'omez et al. 2018b), that the CS evolution is accelerated, low-mass MS progenitors may also explain water masers in young
  PNe as the remnants of the AGB ones, with the only requirement that they are oxygen-rich in the AGB phase.
  Moreover, as already mentioned, WFs have also been associated with 4--8\,M$_{\odot}$ MS progenitors (Section\,4.1). The highest mass range ($\sim$6--8\,M$_{\odot}$)
  may probably be discarded because the extremely fast CS evolution would result in photoionization of envelope in a few or several decades, in consequence, in
  free-free radio continuum emission that is not observed in WFs, except in {\it IRAS}\,15013$-$5754. On the other hand, the lack of free-free emission in WFs
  suggests a relatively stable phase to maintain a (small) population of WFs (see G\'omez et al. 2017). In any case, alike H$_2$O-PNe, the observations
  strongly support that WFs host binary/multiple CSs (Imai 2007; Yung et al. 2011; G\'omez et al. 2015a, 2018b; Orosz et al. 2019) and they could
  represent a phase in the evolution of binary/multiple CSs, that is previous to that of H$_2$O-PNe. 

IRAS18061 suggests a new, additional scenario in which water maser emission is reactivated during the post-AGB evolution through
an LTP/VLTP. Noteworthy, the H$_2$O-PN {\it IRAS}\,17347--3139 also presents dual chemistry (Jim\'enez-Esteban et al. 2006;
Hsia et al. 2016) and Jim\'enez-Esteban et al. already proposed the possibility of a thermal pulse to explain it. If this was the
case, a final thermal pulse might probably be discarded because the IRAS fluxes of {\it IRAS}\,17347--3139 are incompatible with
those of IR-[WC] stars (Section\,4.3); an LTP or VLTP would be more appropriate. Furthermore, although H$_2$O-PNe share several common
characteristics (Miranda et al. 2010; G\'omez et al. 2018a), the similarities between IRAS18061 and {\it IRAS}\,17347--3139 are
extraordinary and many properties of IRAS18061 are replicated in {\it IRAS}\,17347--3139 that shows narrow-waist bipolar lobes ending
in bow-shock-like structures at different distance from the centre; some point-symmetric structures; a compact ionised ring
with a high electron density (1--3$\times$10$^6$\,cm$^{-3}$); and water masers distributed along the ionised torus, that most probably trace
an oxygen-rich neutral ring surrounding the ionised one (de Gregorio-Monsalvo et al. 2004; Sahai et al. 2007; Tafoya et al. 2009; Lagadec et al. 2011).
These remarkable similarities strongly suggest very similar formation processes in both PNe. Identifying the spectral type of the central
star of {\it IRAS}\,17347--3139 is crucial to establish whether this PN is related to a born-again episode. In addition, a born-again scenario 
should be investigated for other H$_2$O-PNe.

\section{Conclusions}

We have analysed new optical narrow- and broad-band images, intermediate- and high-resolution long-slit spectra of the H$_2$O-PN
{\it IRAS}\,18061--2505. We have also included in our analysis archival POSS images obtained in $\sim$1951.6 and $\sim$1996.7.

The images show a pinched-waist bipolar PN (size $\sim$40\,arcsec) consisting of narrow-waist bipolar lobes with some point-symmetric
structures, a bow-shock-like structure at the end of each lobe, and a compact (size $\la$0.7\,arcsec) inner nebula at the centre of the object
that is recognisable in all images, except in the POSS\,I-Blue one from $\sim$1951.6. 

The nebular kinematics shows that the bipolar lobes are strongly disrupted and present very different properties from each other. The bow-shock-like
structures exhibit the characteristics of highly collimated outflows/bullets, their properties differ from each other, and are not aligned
with the axis of the bipolar lobes. In the inner nebula, we identify five (spatially unresolved) components in the spectra by their kinematic and
emission properties, which are compatible with a structure consisting in an inner ring, inner bipolar lobes
and high velocity outflows. In addition, the morphokinematic properties of the bipolar lobes and the water masers provide compelling arguments
to conclude that the water masers and, probably, the silicate emission from IRAS18061, arise from a neutral oxygen-rich ring that traces the
dense equatorial region of the bipolar lobes. The bow-shocks are the oldest structure identified in IRAS18061 ($\sim$950--1000\,yr), followed by the
bipolar lobes ($\sim$760\,yr), while the inner nebula is extremely young ($\sim$70--140\,yr). Each of these components presents its own
centroid/systemic (LSR) radial velocity, which may be explained if the CS is a binary/multiple system.

We carried out a spatially resolved analysis of the emission lines, physical conditions and chemical abundances in IRAS18061. The nebula presents a
very low-excitation, electron temperatures of $\sim$8500--17000\,K, and moderate electron densities of $\sim$1400--4600\,cm$^{-3}$, although
we identify a high electron density region ($\sim$2--5$\times$10$^5$\,cm$^{-3}$) in the inner nebula. [N\,{\sc ii}] and [S\,{\sc ii}] emission lines 
are relatively strong, suggesting a shock-excitation mechanism. However, comparison with recent photoionization models and methods to discriminate
between photoionised and shocks in PNe suggest that the spectra from the bipolar lobes may be explained with photoionization alone if the mass of the MS 
progenitor star of IRAS18061 is $\ga$3\,M$_{\odot}$. A comparison of the chemical abundances in the bipolar lobes with models for stellar yields
suggests a mass of $\sim$3--4\,M$_{\odot}$ for the MS progenitor, in agreement with the expectations that H$_2$O-PNe evolve from
intermediate-mass MS progenitors. Nevertheless, the role of shocks in the nebular excitation of PNe is a very complex, still unsolved problem and
the possibility of a low-mass MS progenitor cannot be definitively ruled out. 

The properties of the bipolar lobes and oxygen-rich ring are consistent with those expected from CEE, if several bipolar and non-bipolar collimated
outflows at different directions have been ejected during or after CEE, that have strongly distorted the bipolar lobes. The pinched-waist morphology
indicates a very high density contrast in the CE, suggesting a relatively low mass ratio between central star and the presumable companion. The
bipolar outflows associated to the bow-shocks could be associated with the formation of an accretion disk around a companion
through Roche-lobe overflow or grazing envelope evolution before the CEE.

The inner nebula may be attributed to an LTP or VLTP. Taken into account the age of the bipolar lobes and that CEE may drastically reduced the
duration of the early AGB phase, a VLTP scenario would be favoured if the progenitor mass is $\sim$3--4\,M$_{\odot}$, whereas an LTP would be
favoured for progenitor masses $\la$1.5\,M$_{\odot}$. The non-detection of the inner nebula in the POSS\,I-Blue image
from $\sim$1951.6 indicates that the thermal pulse occurred sometime before that date, while its detection in a spectrum from 1992.4 suggests
that the CS is reheating. Shock excitation exists in the inner nebula, as revealed by the larger
expansion velocities in H$\alpha$ than in [N\,{\sc ii}], and by its inverted ionisation structure, being the size in [O\,{\sc iii}]
larger than in [S\,{\sc ii}]. 

Bipolar lobes, ring, and collimated outflows are identified in both the main and inner shell of IRAS18061, in all these cases suggesting a binary/multiple
  CS scenario. CEE is a plausible possibility to explain the formation of the main shell (bipolar lobes and oxygen-rich ring). We speculate that the formation
  of the inner shell could be related to a second CE phase that occurred when the CS expanded in its return to the AGB after the thermal pulse, and engulfed the
  presumable close companion resulting from the first CEE phase.

The survival time of water masers after the AGB and the age of the bipolar lobes are incompatible with the presence of water maser emission in IRAS18061.
We propose that the water maser emission in IRAS18061 has been lately reactivated in the post-AGB phase, through shocks propagating in the pre-existing
oxygen-rich ring, that are generated by the thermal pulse. This implies that the CSs of IRAS18061 is not at the very moment of its first entrance in
the PN phase, in consequence, IRAS18061 cannot be considered as an extremely young PN. The distinction between LTP and VLTP could strongly depend on
the MS progenitor mass-

We discussed H$_2$O-PNe and WFs in the light of the results obtained for IRAS18061. Although other mechanisms (e.g., jets, outflow-envelope interaction) may
  excite water masers in the very early PN phase, if CEE is usually involved in the formation of H$_2$O-PNe, accelerating the CS evolution, progenitor masses
  of $\sim$3--4 as well as $\la$1.5\,M$_{\odot}$ may explain water masers in PNe as the remnants of the AGB ones. The properties of WFs strongly support the idea
  that they are also associated with binary/multiple CSs and could represent an evolutionary stage that is previous to that of H$_2$O-PNe.

\section*{Acknowledgments}
This paper is based on data obtained (1) with ALFOSC, which is provided by the Instituto de Astrof\'{\i}sica de Andaluc\'{\i}a (CSIC) under a joint
agreement with the University of Copenhagen and NOTSA (Nordic Optical Telescope Scientific Association); (2) at the Nordic Optical Telescope, operated
by the NOTSA at the Observatorio del Roque de los Muchachos, La Palma, Spain, of the Instituto de Astrof\'{\i}sica de Canarias; (3) at Centro Astron\'omico
Hispano Alem\'an (CAHA) at Calar Alto operated jointly by Instituto de Astrof\'{\i}sica de Andaluc\'{\i}a (CSIC) and Max Planck Institut f\"ur Astronomie
(MPG). Centro Astron\'omico Hispano en Andaluc\'{\i}a is now operated by Instituto de Astrof\'{\i}sica de Andaluc\'{\i}a and Junta de Andaluc\'{\i}a; (4)
in the Observatorio Astron\'omico Nacional at Sierra de San Pedro M\'artir (OAN-SPM) operated by the Instituto de Astronom\'{\i}a of the
Universidad Nacional Aut\'onoma de M\'exico; (5) with the Aristarchos telescope operated on Helmos Observatory by the Institute for Astronomy, Astrophysics,
Space Applications and Remote Sensing of the National Observatory of Athens. We thank Calar Alto Observatory for allocation of director's discretionary time to
this programme. We are very grateful to the staff of Calar Alto for carrying out these observations.
We acknowledge the Instituto de Astrof\'{\i}sica de Canarias for the use of the IAC filters at the NOT. We are very grateful to the staff of all
four observatories for helping during the observations, and to G. Melgoza-Kenedy for those at the OAN-SPM. LFM and JFG 
are partially supported by MCIU grant AYA2017-84390-C2-1-R, co-funded by FEDER funds and from the State Agency for
Research of the Spanish MCIU through the "Center of Excellence Severo Ochoa" award 
for the Instituto de Astrof\'{\i}sica de Andaluc\'{\i}a (SEV-2017-0709). Part of this work has been carried out during stays of LFM at the
Observatoire de la C\^ote d'Azur (Nice, France) and at ALMA Central Office (Santiago, Chile). He acknowledges support from the visitor program of
the Observatoire de la C\^ote d'Azur and from the Joint ALMA Observatory visitor program. RV,
LS, and PFG were supported by UNAM-PAPIIT grants IN106720 and IN101819. LU acknowledges support from grant PE9-1160 of
the Greek General Secretariat for Research and Technology in the framework of the program Support of
Postdoctoral Researchers (Greece). AA acknowledges support from Government of Comunidad Aut\'onoma de Madrid (Spain) through
postdoctoral grant `Atracci\'on de Talento Investigador' 2018-T2/TIC-11697. This research has been partly funded by the Spanish State Research Agency (AEI)
Project MDM-2017-0737 at Centro de Astrobiolog\'{\i}a (CSIC-INTA), Unidad de Excelencia Mar\'{\i}a de Maeztu. IdG is partially supported by MCIU-AEI (Spain) grant
AYA2017-84390-C2-R (co-funded by FEDER). AM is supported by MCIU-AEI (Spain) grant AYA2017-88254-P. This research has
made use of the HASH PN database
at hashpn.space, the SIMBAD database, operated at CDS, Strasbourg, France, and of the VizieR catalogue access tool, CDS,
Strasbourg, France (DOI: 10.26093/cds/vizier). The original description of the VizieR service was published in
Ochsenbein, Bauer \& Marcout (2000). The Pan-STARRS1 Surveys (PS1) and the PS1 public science archive have been made possible
through contributions by the Institute for Astronomy, the University of Hawaii, the Pan-STARRS Project Office, the Max-Planck Society and its
participating institutes, the Max Planck Institute for Astronomy, Heidelberg and the Max Planck Institute for Extraterrestrial Physics, Garching,
The Johns Hopkins University, Durham University, the University of Edinburgh, the Queen's University Belfast, the Harvard-Smithsonian Center
for Astrophysics, the Las Cumbres Observatory Global Telescope Network Incorporated, the National Central University of Taiwan, the Space
Telescope Science Institute, the National Aeronautics and Space Administration under Grant No. NNX08AR22G issued through the Planetary Science
Division of the NASA Science Mission Directorate, the National Science Foundation Grant No. AST-1238877, the University of Maryland, Eotvos
Lorand University (ELTE), the Los Alamos National Laboratory, and the Gordon and Betty Moore Foundation. The Digitized Sky Surveys were produced
at the Space Telescope Science Institute (STScI) under US Governmentgrant NAG W-2166. The images of
these surveys are based on photographic data obtained using the Oschin Schmidt Telescope on Palomar Mountain and the UK Schmidt
Telescope. The plates were processed into the present compressed digital form with the permission of these institutions. The National
Geographic Society – Palomar Observatory Sky Atlas (POSS-I) was made by the California Institute of Technology with grants from the
National Geographic Society. The second Palomar Observatory Sky Atlas (POSS-II) was made by the California Institute of Technology with
funds from the National Science Foundation, the National Geographic Society, the Sloan Foundation, the Samuel Oschin Foundation and the
Eastman Kodak Corporation. The Oschin Schmidt Telescope is operated by the California Institute of Technology and Palomar Observatory.
The UK Schmidt Telescope was operated by the Royal Observatory, Edinburgh, with funding from the UK Science and Engineering Research
Council (later the UK Particle Physics and Astronomy Research Council), until 1988 June, and thereafter by the Anglo-Australian Observatory.
Supplemental funding for sky survey work at the STScI is provided by the European Southern Observatory.

\section*{Data Availability}

The CAHA, VLA, and POSS data used in this paper can be accessed through the corresponding archives: http://caha.sdc.cab.inta-csic.es/calto/,
https://science.nrao.edu/observing/data-archive, https://archive.stsci.edu/dss/, respectively. The rest of the data may be obtained upon justified
request to the first author.

\newpage

\clearpage

\appendix

\section{Additional figure: the innermost regions of $IRAS$\,18061--2505}

\begin{figure*}
\begin{center}
   \includegraphics[width=120mm, angle=270]{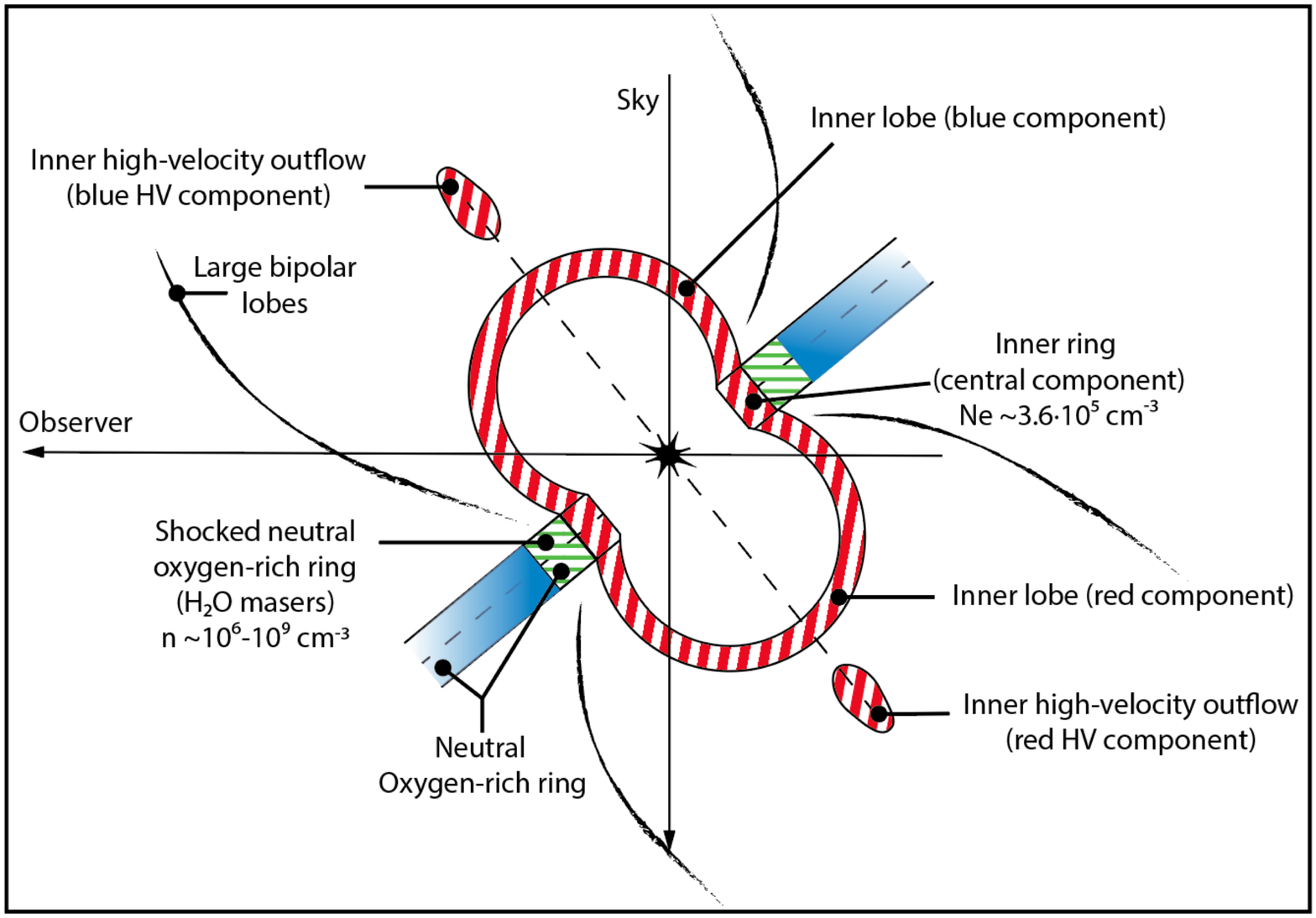}
\caption{Sketch of the innermost regions of IRAS18061. The structures are not to scale. The
  inclination of the inner shell accounts for the redshifted component being fainter than the blueshifted one (Sections\,3.2.1 and 4.2).
  The different structures and regions are labelled, and information about chemistry, particle density ($n$), and electron density ($N$$_{\rm e}$) is provided
  for some of them. The inner shell is probably carbon-rich, as expected from an LTP/VLTP.}
\end{center}
\end{figure*}

\section{The distance of $IRAS$\,18061--2505}


IRAS18061 has been considered as a bulge PN due to its Galactic coordinates
($l$ = $5\rlap.^{\circ}9747$, $b$ = $-2\rlap.^{\circ}6125$). However, most probably this is not the
case because, as noticed by G\'orny et al. (2009), the spectral type of the central
star is not typical for bulge PNe, and the angular size of the nebula ($\sim$40\,arcsec) is much larger
than the upper limit of 20\,arcsec imposed to bulge PNe (Stasi\'nska \& Tylenda 1994). 

We have used the {\it Gaia Early Data Release 3} (GEDR3; Gaia Collaboration, A.G.A. Brown et al. 2020) to search
for a possible parallax of IRAS18061, and downloaded the images of the Pan-STARRs1 survey around the object in the z, g, and r filters
(Chambers et al. 2016; Flewelling et al. 2016). Figure\,A1 shows the field around IRAS18061 as it appears in Aladin-Lite, superimposed
with the sources identified in GEDR3. The core of IRAS18061 is clearly identified and GEDR3 provides the following information for it: \\

\noindent {\it Source ID} = 4065774303370565376 \\
$\alpha$(2016.0) = $18^{\rm h}$ $09^{\rm m}$ $12\rlap.^{\rm s}411$ \\ 
$\delta$(2016.0) = $-25^{\circ}$ 04$'$ $34\rlap.^{''}56$ \\
G magnitude = 16.966175$\pm$0.0013412\,mag \\ 
Parallax = $-$0.5450$\pm$0.4355\,mas \\

The negative Parallax for the core of IRAS18061 is most probably due to its nebulous nature
and, in consequence, the difficulty of measuring its photocenter.

Distance estimates to IRAS18061 using statistical methods are 1.3\,kpc (Preite-Mart\'{\i}nez 1988), 2.8\,kpc 
(Tajitsu \& Tamura 1998), and 6.62\,kpc (Vickers et al. 2015). The third distance may be ruled out because it has
been obtained assuming that IRAS18061 is a bulge PN and, in fact, places the object in the bulge (see above).
In this paper, we will adopt a distance for IRAS18061 of 2\,kpc, as the mean value of 1.3 and 2.8\,kpc. A more precise
value for the distance may be obtained from measurements of the parallax of its water masers using VLBI techniques, as in
the case of the H$_2$O-PN K\,3-35 (see Tafoya et al. 2011).

We note that a search of IRAS18061 in Vizier provides a parallax from GEDR3 of 0.5534$\pm$0.1214\,mas
for an object at 1.414\,arcsec from the position of the core of IRAS18061. Because no (bright) star is observed
within a radius $<$1.5\,arcsec from the core, it is tantalizing to associate that parallax to the
core. However, a detailed inspection of the data and images show that this is not the case and that the parallax
refers to the field star arrowed in Figure\,A1, for which GEDR3 provides: \\

\noindent {\it Source ID} = 4065774307705264384 \\
$\alpha$(2015.5) = $18^{\rm h}$ $09^{\rm m}$ $12\rlap.^{\rm s}965$ \\ 
$\delta$(2015.5) = $-25^{\circ}$ 04$'$ $34\rlap.^{''}03$ \\
G magnitude = 17.904875$\pm$0.003966\,mag \\ 
Parallax = 0.5534$\pm$0.1214\,mas \\

This discrepancy may be due to errors in the coordinates of IRAS18061 in SIMBAD (see Figure\,A1, green square) or
confusion of that coordinates with those of the field star marked in Figure\,A1.

\begin{figure}
\begin{center}
\includegraphics[width=81mm]{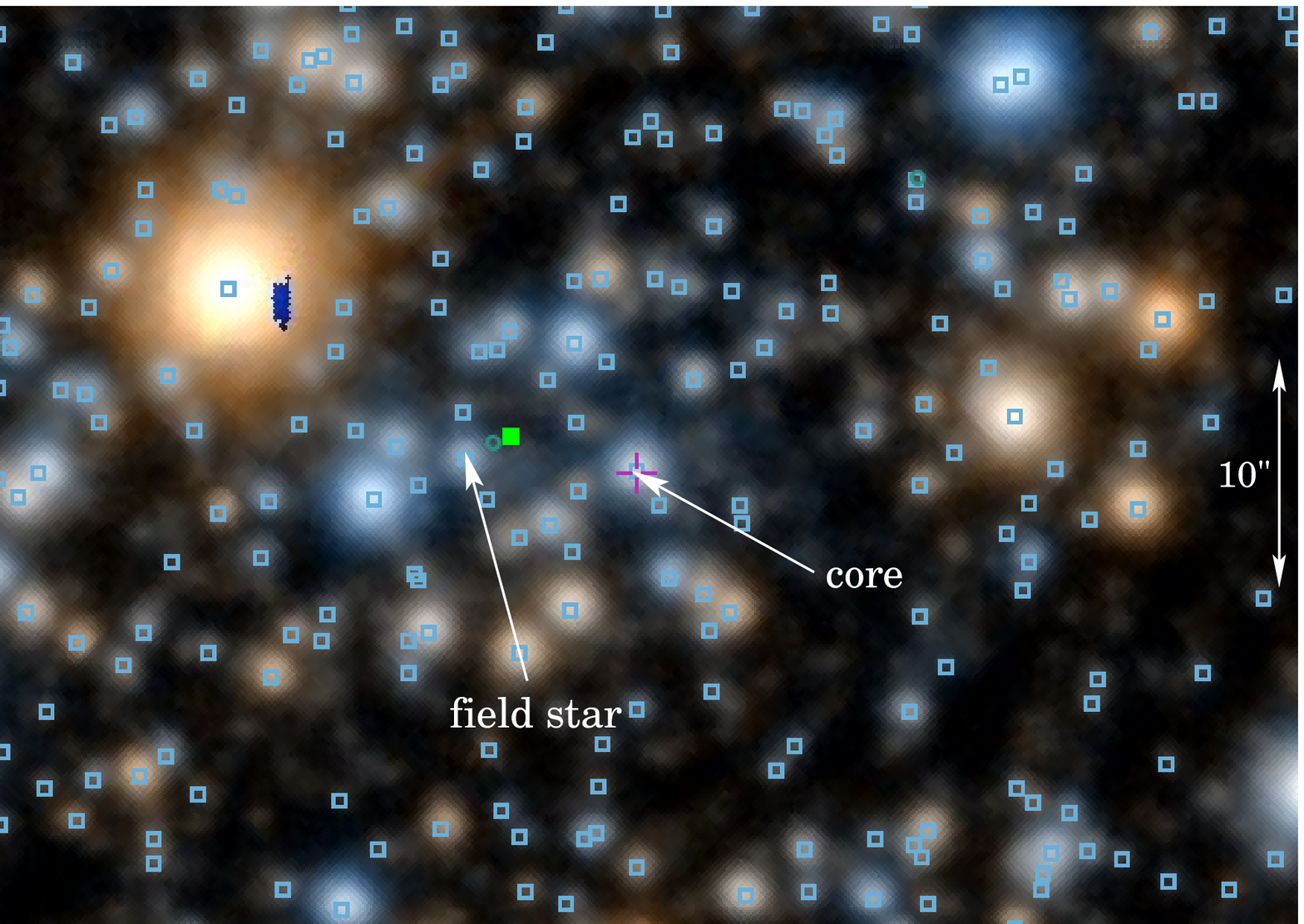}
\caption{Colour composite image of the field around IRAS18061 obtained by combining the images in the z (red) and g (blue) filters
  from the PanSTARRS archive. The core of IRAS18061 and the field star discussed in the text are arrowed, the empty blue squares mark the
  sources identified in GEDR3, and the filled green square marks the position of IRAS18061 from SIMBAD. North is up, east to the left,
  and the angular scale is indicated. }
\end{center}
\end{figure}

\section{SHAPE reconstruction of the bipolar lobes of {\it IRAS}\,18061$-$2505}

We have used the tool {\sc shape} (Stephen et al. 2011) to reconstruct the morphokinematic structure of IRAS18061, based on
the images and high-resolution long-slit spectra. We have concentrated on the extended emission from the bipolar lobes
as observed in the [N\,{\sc ii}] image and in the PV maps at PAs +33$^{\circ}$, +55$^{\circ}$, and +60$^{\circ}$. The spectrum at
PA $-$81$^{\circ}$ will not be considered because the kinematics at this PA is not typical of a bipolar shell but of the action of
a collimated outflow on the shell (Section\,3.2.2). The multiple knots observed in the PV maps will not be
considered because no constrains can be imposed on the parameters of ``isolated'' knots (see Section\,3.2.2). We have
followed the standard process to reconstruct a bipolar nebula, starting with a spherical structure that is then deformed by the
modifiers {\it Squeeze} and {\it Bump}, in a self-consistent way, so that image and PV maps must be simultaneously reproduced.

\begin{figure*}
\begin{center}
\includegraphics[width=175mm]{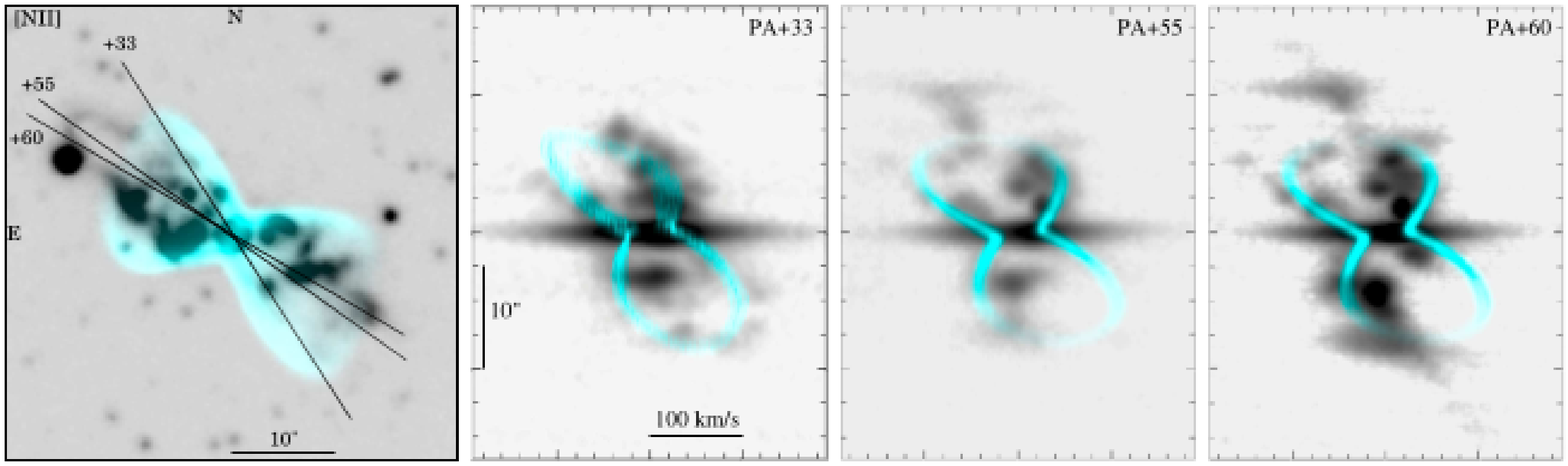}
\caption{Shape reconstruction of the bipolar lobes of IRAS18061. The reconstructed image and PV maps are shown in cyan, superimposed
on the data. The slit positions of the PV maps are schematically drawn on the image for reference (see Figure\,5 for more details).}
\end{center}
\end{figure*}

The reconstructed image and PV maps are shown in Figure\,C1 (cyan) superimposed on the observed data. The basic structure of the bipolar lobes
is well reproduced, although some details are not well addressed. For instance, some faint features at PA +55$^{\circ}$ and +60$^{\circ}$,
that appear separated from the shell, and, particularly, features observed at PA +33$^{\circ}$, that deviated from
the general tendence. These differences may be attributed to peculiar local motions that are caused by jet--shell
interaction. 

The best fit model was obtained iteratively and corresponds to a distorted bipolar shell with main bipolar axis at PA=+63$^{\circ}$ (very
similar to that deduced from the images), inclination angle respect to the plane of the sky $i$=15$^{\circ}$, and a homologous velocity law
$V$[km\,s$^{-1}$]=12\,r[arcsec]. The polar radius and expansion velocity are 13\,arcsec and 163\,km\,s$^{-1}$, respectively, implying a kinematical age
of $\sim$760\,yr with an estimated error of $\pm$60\,km\,s$^{-1}$. Nevertheless, expansion velocities of 190--200\,km\,s$^{-1}$ are required along
PA +33$^{\circ}$, which are consistent with acceleration of these regions by the impact of collimated bipolar outflow on the shell.

\bsp	
\label{lastpage}
\end{document}